# The Reasoning Under Uncertainty Trap: A Structural AI Risk

*Toby D. Pilditch*

*Transformative Futures Institute,*

*University of Oxford, and University College London*

**Abstract**

This report examines a novel risk associated with current (and projected) AI tools. Making effective decisions about future actions requires us to reason under uncertainty (RUU), and doing so is essential to many critical real world problems. Overfaced by this challenge, there is growing demand for AI tools like LLMs to assist decision-makers. Having evidenced this demand and the incentives behind it, we expose a growing risk: we 1) do not currently sufficiently understand LLM capabilities in this regard, and 2) have no guarantees of performance given fundamental computational explosiveness and deep uncertainty constraints on accuracy. This report provides an exposition of what makes RUU so challenging for both humans and machines, and relates these difficulties to prospective AI timelines and capabilities. Having established this current potential misuse risk, we go on to expose how this seemingly additive risk (more misuse additively contributed to potential harm) in fact has multiplicative properties. Specifically, we detail how this misuse risk connects to a wider network of underlying structural risks (e.g., shifting incentives, limited transparency, and feedback loops) to produce non-linear harms. We go on to provide a solutions roadmap that targets multiple leverage points in the structure of the problem. This includes recommendations for all involved actors (prospective users, developers, and policy-makers) and enfolds insights from areas including Decision-making Under Deep Uncertainty and complex systems theory. We argue this report serves not only to raise awareness (and subsequently mitigate/correct) of a current, novel AI risk, but also awareness of the underlying **class** of structural risks by illustrating how their interconnected nature poses twin-dangers of camouflaging their presence, whilst amplifying their potential effects.

# Executive Summary

Effective determination of prospective actions requires us to reason from incomplete, imperfect information towards uncertain futures. This reasoning under uncertainty (RUU), though difficult, is essential to avoiding unnecessary harms, and maximising potential gains. The larger (and often more complex) the problems we seek to solve, the greater the challenge, and the more there is to gain (or lose). Here we describe a current and growing structural AI safety risk that extends from this principle. In the first three chapters, we build our case from three core premises, providing evidence to support each.

In chapter one, we lay out the nature of reasoning under uncertainty for a general audience, summarising the current known status of human performance on these tasks, and how they relate to the problems we face. From this we derive the first premise: RUU is hard for humans, and particularly when dealing with the wicked problems we face on larger scales (e.g., government, industry, military).

In chapter two, we pivot to the organisational level, outlining the recent history of demand for assistance in dealing with the challenging problems decision-makers face, and outline the wider spectrum of incentives and considerations organisations (be they private or public sector) consider when not only seeking, but integrating new techniques, tools, training, and solutions. Finally, we provide recent evidence from multiple sectors across the globe of a growing demand to use recent AI advances (namely LLMs) to assist organisations (be they industry heads, policy-makers, and more) with the difficult tasks of reasoning under uncertainty (e.g., policy and strategy planning). This sets us the second premise: There is growing demand among powerful decision-makers / organisations to deploy LLMs for RUU tasks, given the challenges they face, and the incentives at play.

In chapter three, we provide an exposition of why RUU is so challenging, such that the reader may understand the difficulties faced not only by human reasoners, but by AI tools. Placing these processes within the context of the hard (*wicked*) problems organisations are attempting to reason about, we explain the two fundamental difficulties of RUU: combinatorial explosiveness (across decision spaces, information sources, and explanatory models) and ***deep*** uncertainties. Relating this to current and prospective AI tool capabilities (up towards AGI/ASI), we set out the final premise: The AI tools (i.e., LLMs) that organisations are turning not only have little positive evidence for RUU competence but may in fact face fundamental challenges if applied in this manner. Critically, there is a lack of awareness of these potential shortcomings among prospective users (and arguably insufficient attention across the development space). We lastly distinguish between AI RUU optimal performance (which faces fundamental performance ceilings), better-than-human performance (of which there is no current positive evidence), and relate these two prospects to AI development timelines.

Having set out these three premises, we outline the central risk highlighted in this work: a current and growing risk of AI tool RUU misuse within organisation decision-making (and beyond), and particularly when tackling larger, more difficult (and important) problems - stemming from a general lack of awareness of performance limitations, and systemic use/uptake incentives. Chapter four builds upon this central problem, by placing it within the larger context of the structural risks that induce it, and in so doing introducing the "Trap" dynamic of the RUUT. Specifically, we outline how system behaviours and relationships produce a series of compounding feedback loops that turn a seemingly linear rate of misuse error costs into an accelerative, non-linear one. For example, we illustrate how problem opacity not only spurs demand, but prevents easy detection of errors, fostering misplaced trust, weakening oversight incentives, which feedback into further (mis)use and establishment of (unknowingly) harmful norms, which in turn signal further misplaced trust and widespread uptake. We further outline how the impacts of these erroneous (and costly) decisions can be catastrophic in the wider, global problem space.

We then relate RUUT to the wider context of structural risks, and highlight several AI risk topics with common features and linkages in chapter five, including long-horizon planning capability detection, and automation of science vulnerabilities.

In chapter six we turn to potential solutions for mitigating the RUUT, not only as a specific risk, but also as emblematic of the solutions/perspective required for tackling socio-technical / structural risks. In so doing, we outline a three-stage process of awareness, (formal) investigation, and intervention components, addressing not only tool development and use directly, but also surrounding AI safety policy. Our proposals are savvy to the structural risk aspect of the problem (i.e., leaning into, rather than working against, the positive feedback loops in play), and integrate insights from the Decision-making under uncertainty (DMDU) and complex systems literatures (e.g., shifting performance metrics and prompts based on identified problem features/dimensions, and using multi-lever interventions that target sensitive intervention points in the wider system).

Finally, in chapter seven we conclude by setting out the primary goals of this work:

First, we seek to raise awareness of a potentially costly structural risk associated with the (mis)use of AI tools, the RUUT, such that we may best avoid misuse and promote the design of effective solutions. In so doing, we also seek to emphasise that this particular risk does not require *optimal* RUU performance from AI tools to resolve, it requires accurate awareness of current RUU performance across applied problem areas, such that any improvements in RUU performance beyond human standard performance (something we believe is eminently feasible, especially with integration of hard-won insights from fields such as DMDU) can start to instead yield net positive outcomes.

Second, we aim to raise awareness for this **class** of risk within the AI safety and governance space - structural risks (gelling with recent calls for socio-technical perspectives). We show <u>how an initially "small" impact risk can compound to outsized effects</u> when one considers the feedback and compounding impact of systems and structure. This raises grave implications and timing imperatives, as impacts can snowball rapidly. Further, we emphasise how both technical and policy solutions should account for these risks in their design. Specifically, as we outline here, structural risks can be value neutral accelerative forces (e.g., widespread uptake), and as such can be harnessed to yield positive outcomes, akin to leverage points in complex systems. Effective AI safety (and wider) policy, given rapidly evolving and uncertain timelines and consequences will need to integrate this perspective.

# Chapter 1: Human Reasoning in an Uncertain World

## 1.1. Reasoning Under Uncertainty Introduction

Navigating our way in the world requires reasoning. To make an effective decision requires us to use the information available to us so as to best estimate possible futures. This could be as simple as determining what to eat for lunch ("this sandwich tasted good previously, so I will eat it again"), to planning our futures ("where should I buy a house?"). To arrive at a conclusion requires us to *reason* (i.e., make inferences) about possible effects our decision/action might have, and thus determine its worth.

Reasoning entails two processes:

**Selection**: First, we must *select* relevant information, be this prior experiences, past feedback, situational cues, or proffered advice.

**Integration**: Second, we need to synthesise or *integrate* this information into a coherent picture, such that we may *infer* the appropriate conclusion.

These two processes are essential to not only reasoning about the future (predictive reasoning), but also about past events (diagnostic reasoning). The degree to which we engage in these processes will vary from decision to decision. I am far more inclined to gather a lot of information when considering purchasing a house, than I am when buying a sandwich. Of course, these processes are *effortful*; it takes time and cognitive load to both select a sufficient amount of information, and consequently process and integrate that information into a coherent whole. As a result, the degree to which we engage in these processes tends to be correlated with the size of the prospective outcomes - getting a house purchase wrong is more costly than buying a so-so sandwich. I may spend months gathering information for the former, but I am not going to waste more than a few minutes considering my recent history of sandwich purchases for the latter.

Regardless of how invested we are in arriving at an accurate conclusion, these processes are never perfectly accurate. We are typically bereft of the complete picture when needing to make a decision. There is always *some degree of uncertainty*. Every item of evidence or information we consider can have uncertainty stemming from several different sources:

- **Incompleteness**. Missing or incomplete data.
- **Imprecision**. Measurement error.
- **Unreliability**. A sources (lack of) expertise and/or trustworthiness.

These sources of uncertainty are not mutually exclusive (not only have I not tasted every previous sandwich, but my sense of when a sandwich has been tasty is not entirely reliable), and in fact can compound (e.g., I am uncertain/unreliable in knowing how reliable or unreliable my sense of taste was for each previous sandwich). Regardless of the degree or combination, suffice to say when we are trying to reason about the future and possible actions, there will *always* be some uncertainty. Hence, *reasoning under uncertainty* (RUU).

At this juncture, it is important to flag three running themes that we will return to throughout this work:

1. **Selectivity**. That *good reasoning* requires good information selection, and that this includes prior experience. Critically, this includes identified prior mistakes.

2. **Uncertainty**. That there are multiple sources of uncertainty, and any number of these may be irresolvable on any given prediction/decision (see radical/deep uncertainty, e.g., Keynes, 1921). Furthermore, these uncertainties can compound and combine.
3. **Integration difficulty**. With each additional item of information, the harder the integration challenge. As we will touch on later, this is not just due to the difficulty of finding the coherence across information items (e.g., when data conflicts), but due to the compounding uncertainties applying not just to each new information item, but to the interactions (or relations) between an information item and *every other information item*. This is also known as information *fusion* (for a related list of challenges in this theme, see Davis et al., 2016).

These considerations are "in play" regardless of the magnitude of the reasoning endeavour, from everyday decisions to longer term planning. However, and central to this paper, not only must individuals make decisions that entail reasoning about an uncertain future, but organisations and institutions must similarly make decisions and formulate plans to achieve their goals. Instead of a personal consideration of where one should buy a house, or purchase a sandwich, these might be decisions about national housing policy, or whether it is worthwhile to continue stocking sandwiches at all. Before we address these more substantive, harder reasoning problems, outlining how they complicate an already challenging cognitive process, we will first address the current understanding of human RUU, against which an AI comparison will be judged.

*1.2. Human RUU weakness*

It is necessary to clarify how effective humans are at RUU if we are to get a sense of how overfaced we may be in the face of current and future reasoning problems. To do this, we turn to the field of cognitive psychology which, among other domains, has empirically investigated the strengths and weaknesses of human RUU.

Empirical work, wherein participants are providing reasoning (or markers of reasoning, such as decisions and predictions) is conducted under *controlled conditions*. Specifically, there is a constrained goal or question that the reasoning should address, and there is a defining scope of enquiry that (intends to) limit the possible variables (or items of evidence/information) necessary to form a conclusion.

*1.2.1. Axiomatic Comparison*

Empirical work in cognitive psychology has investigated reasoning under uncertainty through the lens of probabilistic reasoning and an axiomatic basis, which posits that uncertainties can be expressed through probability distributions. This assumption, strong and sometimes controversial, underpins much of the scientific inquiry into how humans confront uncertain scenarios.

Probabilistic reasoning relies on the notion that individuals can and should utilise probability to guide their beliefs and decisions when confronted with uncertainty. The axiomatic framework further proposes that rational decision-making is achieved by minimising inaccuracy and making optimal inferences, often measured against a Bayesian standard. This rigorous structure provides a benchmark for assessing human reasoning – but it is also predicated on assumptions that may not be reflective of real-world conditions. The historical study of RUU often involved controlled tasks with known probabilities, such as drawing balls from urns or selecting cards from a deck, or more recently comparison to specific vignettes or scenarios with pre-defined probabilities and underpinning problem structures. These tasks were structured to yield precise probabilistic conclusions, enabling a clear assessment of reasoning accuracy.

This research revealed that human performance frequently deviates from the axiomatic ideal. More broadly, biases and errors are considered to stem from a handful of parent causes. These include (but are not limited to) errors of anchoring (early information biases subsequent search or integration; Pitz, 1969; Lord et al., 1979), availability (skewed sampling from memory, Tversky & Kahneman, 1974), confirmation (seeking to confirm a preconception/belief, Macdougall, 1906; Nickerson, 1998), ego (too heavy reliance on own perspective, Ross et al., 1977), framing (susceptibility to the presentation, rather than content of information, Tversky & Kahneman, 1981), logic and risk (more direct violations of logical and axiomatic principles, e.g., conjunctions; Tversky & Kahneman, 1983), and errors of scoping/sample (neglecting relevant events or components, such as sample size or timeframe; Jarvik, 1951; Gilovich et al., 1985).

Within the context of RUU, it is useful to consider errors and biases in two categories: the *selection* of evidence/information, and the *integration* of that evidence into a coherent whole. This is useful for two reasons: First, from a psychological perspective, it allows one to relate types of error to the point in the process - critical to determining possible corrections. Second, if we are to later consider AI tool strengths and weaknesses, it is useful to determine capabilities along the lines of sufficient information gathering and selection capabilities, and the integration or computation of the information available (e.g., is an issue attributable to lack of information, or a poor integration algorithm).

Some examples of selection and search biases and errors include confirmation bias (seeking information that confirms one's preconceptions; Macdougall, 1906), positive search strategies and selection biases (seeking information that confirms a favoured hypothesis/rule, rather than falsify it; Wason, 1960). The commonality to errors in this part of the reasoning process is not necessarily that the information chosen is *wrong*, but rather that it is *insufficient* to effectively inform the reasoning processes that follow.

Errors in the integration of evidence include confirmation bias once again (in this case, overweighting confirmatory information, Nickerson, 1998), inertia (resistance to belief adjustment, Phillips & Edwards, 1966), underestimation of multiply-explained evidence (Pilditch et al., 2019), both over and under estimation of the impact of dependencies among items of evidence (Pilditch et al., 2020), and insufficient correction in the face of contradictory evidence (Desai et al., 2020). The commonality to errors in integration is even with sufficient information inputs, reasoning outputs deviate from normative expectation due to systematic over or under-weighting of those inputs - often when that correct normative expectation is more cognitively burdensome.

These studies indicate that when it comes to both selecting and integrating evidence, humans often fall short of axiomatic rationality. We are prone to a range of biases and heuristics that skew our reasoning, leading to systematic errors.

This dim perspective on human reasoning performance should, however, be contextualised within the constraints of axiomatic systems themselves. The adherence to axiomatic benchmarks implies a belief that all uncertainties are tractable and reducible to structured tasks – a notion not universally accepted. Critics argue that real-world uncertainties are often complex and multidimensional, resisting neat probabilistic encapsulation. Moreover, axiomatic rationality assumes a level of cognitive computation and statistical literacy that may not be reasonable for the average person. Thus, while the pursuit of rationality underpins much of cognitive psychology's research into RUU, the benchmarks against which we measure human reasoning are deeply embedded in philosophical and mathematical presuppositions that may not align with everyday cognition.

In essence, the axiomatic perspective, with its strong assumptions, sets a high (and specific) bar for human reasoning – one that may not be entirely fair or representative of the way people actually process uncertainty in their daily lives. Whilst understanding where and why human reasoning diverges from this ideal can provide critical insights into how people think, learn, and make decisions, this is not the only perspective on human reasoning performance.

*1.2.2. Adaptively Rational Reasoning*

Rationality in human RUU has historically been interpreted through the lens of axioms and probabilistic reductions. However, alternative accounts have emerged that challenge this perspective by proposing that human reasoning is not solely an exercise in mathematical computation but is also an adaptive tool honed by evolution and shaped by environmental demands.

Reasoning for the purpose of communication has been highlighted by researchers such as Mercier and Sperber (2011) and furthered by their subsequent works. They suggest that reasoning is primarily used as a means of argumentation to persuade others rather than as a mechanism for discovering truth. While this approach provides fresh insights, it has faced criticism for issues related to falsifiability and the inherently verbal nature of the theories (see Chater & Oaksford, 2018).

Alternatively, adaptive rationality has garnered attention for offering reinterpretations of findings previously understood through an axiomatic framework. This perspective posits that humans exhibit satisfactory accuracy in reasoning when considering the cognitive costs and the typical environmental models they face (for a review, see Hahn and Oaksford, 2007). This approach has led to the reinterpretation of previously (axiomatically) irrational findings, such as the determination of a positive test strategy "error" (Wason, 1960), as rational given the information constraints of the reasoner (Oaksford & Chater, 1994). The adaptive rationality framework urges a broader consideration of the problem contexts and the impacts of experience and learning on reasoning processes. It recognizes that while a probability-based approach may be restrictive, it is often adequate for the environments in which our reasoning capabilities evolved. This underscores the idea that what might have been deemed biases from an axiomatic standpoint can actually be seen as rational adaptations to real-world constraints.

Relatedly, work in causal reasoning has shed new light on previously axiomatically irrational reasoning errors, such as our understanding of base rate neglect (Krynski & Tenenbaum, 2007) - wherein the error diminishes in the presence of a causal framework, suggesting that individuals are better attuned to reasoning about causes than pure statistics. Work in this area has argued that causal schemas serve as mental overlays that integrate probabilistic elements into a broader adaptive reasoning process. This can lead to more effective decision-making through mechanisms like model pruning and selection (Lagnado, 2021; Pearl 1988; 2009).

Axiomatic rationality, though providing a stringent metric for evaluating human reasoning, may be too unforgiving as a basis for judgement. The adaptive capabilities, which some axiomatic perspectives might categorise as biases, have served humans well in typical environments.

Nevertheless, these adaptive strategies are not as effective when tackling complex or substantive problems that go beyond our evolutionary adaptations.

However, it's crucial to recognize that human capacity for reasoning under uncertainty is not static. Research has shown when we are in possession of rich and well-informed priors, we are quite capable (and approximately Bayesian rational) of drawing accurate inferences from those priors (Griffiths & Tenenbaum, 2006). The opportunity to learnHuman reasoners are also capable of learning and calibration. This adaptive learning can lead to surprisingly varied performance in RUU tasks. As we accumulate experience and knowledge, our reasoning processes can become more aligned with the demands of more complex evaluative problems.

*1.2.3. Calibration, superforecasting, and constraints*

The most salient application of human RUU of relevance to the problem at hand is forecasting. As a discipline, it has ancient roots stretching back to the days when augurs would predict the future from the flights of birds or the entrails of animals. In modern times, forecasting has evolved into a more empirical field, with applications ranging from meteorology to finance. In the realm of cognitive psychology, forecasting serves as a critical domain for understanding how individuals reason under uncertainty and how they might improve their judgement over time. In particular, the science of forecasting has grown around statistical and probabilistic methods to determine degrees of belief in future events, and associated forecasting accuracy.

The Good Judgment Project (GJP) marked a significant evolution in forecasting by hosting tournaments designed to test and improve human judgement (Tetlock & Gardner, 2016). These tournaments entailed generating questions that are specific, time-bound, and have clearly verifiable outcomes. Questions like "Will country X enter a recession in the next year?" or "Will the incumbent win the next presidential election in country Y?" are typical, designed to be scored for accuracy post-outcome. In these tournaments, forecasters' predictions are quantified using Brier scores (Brier, 1950)[1], providing a numerical measure of accuracy. Questions are carefully crafted by experts to ensure they are unambiguous and measurable, allowing for precise scoring and feedback. This work has popularised forecasting, as evidenced by the proliferation of forecasting websites and prediction markets, both public (e.g., PredictIt or Metaculus), and internally within organisations.

The term "superforecaster" was coined to describe individuals who consistently outperform their peers in these tournaments. Only a tiny fraction (approximately 1.5% of 14,300 participants from over a decade of Good Judgment research) reach this elite status (Karvetski, 2021). These individuals are characterised by their frequent and meticulous updates to predictions and their superior calibration, achieving Brier scores that are notably better than average forecasters - individually achieving Brier scores of 0.166 on average, as compared to "regular" forecaster performance of 0.259 - notably worse than random guessing. Notably, forecasting accuracy was improved further via aggregation of forecasts across individuals (0.146 in superforecasters, and 0.195 in regular forecasters), illustrating a wisdom of the crowd effect - something that should be noted when considering organisational processes for forecasting.

---

[1] A Brier score is calculated via the squared error between a forecasted probability (between 0 - definitely will not happen, and 1 - definitely will happen) and the forecasted event occurring (1) or not (0). As such, completely incorrect confidence produces a maximum score of 1 ($(0 - 1)^2$, or $(1 - 0)^2$), stating a 50/50 probability produces a random guessing score of 0.25 ($(.5 - 1)^2$), and completely calibrated, perfect accuracy produces a Brier score of 0 (e.g., $(1 - 1)^2$).

Whilst this forecasting accuracy is impressive, there are several important limitations in regards to the general background of human RUU performance - and particularly on substantive problems.

First, superforecasters are notably an outcome-defined minority of a subset of the population (i.e., they are classified as such based on *determined accuracy*, and are a small percentage - 1.5% - of a self-selecting group of interested individuals, Karvetski, 2021). This makes inferring clear reasoning process causes of increased accuracy difficult (Beard et al., 2020), and consequently means the question of how possible it is to train someone to become a superforecaster is unresolved.

Second, and perhaps more critically, forecasting questions are necessarily highly constrained. The calibre of forecasting seen in tournaments is contingent on the nature of the questions posed. The clear-cut, well-defined parameters of tournament questions do not represent the messiness of real-world scenarios. This structured environment is conducive to calibration, but the feedback it provides is of questionable real-world applicability - given the myriad complex, feedback-scarce RUU problems that individuals and organisations face (e.g., no clear time-boundaries, a plurality of possible unknown outcomes, and myriad action/intervention options - to name just a few considerations). Consequently, the narrowly defined questions and the specialised population of forecasters may present a skewed picture of human reasoning abilities.

We turn to these more complex, *wicked* problems next, where we must not only consider the uncertainty inherent to these problems, but also the uncertainty of just how good human RUU performance in such environments can be when bereft of feedback and the opportunity to calibrate.

## *1.3. Wicked Problems and RUU*

In the sphere of high-level decision-making, whether within or across organisations, the ability to reason under uncertainty is not just an academic exercise; it is a critical, ongoing operational necessity. Organisations are tasked with charting courses through uncertain futures, requiring them to make complex multi-criteria decisions that must stand the test of time and unpredictable outcomes. The crux of the issue lies in dealing with multistage strategies, conflicting incentives, collective decision-making processes, time-lagged outcomes, and the extensive temporal spans over which these decisions unfold (on long range prediction difficulty see e.g., Armstrong, 2001; Mellers et al., 2015).

More formally, we can consider these problems as "*wicked*" (Rittel & Webber, 1973). The challenge in dealing with these problems stems from irresolvable/unknowable uncertainties, based on the properties of both the problem space, the necessary goals of the decision-maker, and the complexity of the possible solution space. In particular, the following *wicked* problem features can all have direct bearing on RUU difficulty in isolation, but in combination can prove substantially more detrimental:

- **No Definitive Formulation**: Each wicked problem is a unique nexus of interdependencies and uncertainties, making it impossible to encapsulate in a single, definitive problem statement.
- **No Stopping Rule**: Wicked problems do not have a clear end-point or a defined moment when a solution is reached; they are continuous without a natural conclusion.
- **Solutions as Hypotheses**: Responses to wicked problems are not right or wrong in a binary sense but are better understood as hypotheses subjected to continual testing and refinement.

- **No Ultimate Test**: There is no definitive experiment or moment of truth that confirms the efficacy of a solution to a wicked problem; outcomes are often seen in shades of grey rather than black or white.
- **One-Shot Operations**: Each attempt to solve a wicked problem is significant and has real-world impacts, limiting the ability to learn from trial and error.
- **No Enumerable Set of Solutions**: Wicked problems cannot be exhaustively described in terms of potential solutions; they require creative and innovative approaches.
- **Uniqueness**: Every wicked problem is a one-off. The uniqueness of each scenario means that experiences cannot be fully generalised.
- **Symptoms of Another Problem**: Wicked problems are often entangled with other issues; solving one aspect may exacerbate another.
- **Multiple Explanations**: Discrepancies within wicked problems can be explained in numerous ways, and the selection of a narrative influences the approach to the solution.
- **Responsibility of the Planner**: Social planners and decision-makers carry the burden of the consequences, as there is no right to be wrong.

Let us revisit the housing problem. Consider a policy-maker tasked with improving the housing market. Their reasoning must contend with a continuously changing system (i.e., the market never "stops"), myriad possible explanations for those changes (e.g., economic, socio-political, energy, and other dimensions), and thus a relatively unique scenario at any given time point. Even in determining possible levers (which, when, and in what combination / degree), the policy-maker must also determine a balance between myriad stakeholders (e.g., renters, landowners, developers), making a true determination of "improvement" challenging, and the means of getting there a selection across a wide dimension-space. Lastly, indicators of "improvement" may be similarly mirky, with imperfect measurements that, for example, may suffer from (unknown) changing degrees of validity and accuracy.

This type of deep uncertainty challenge, and the broader wicked problem paradigm applies across myriad domains—social and economic policy, military decision-making, business strategy, and beyond. These problems pervade high-level strategic thinking in almost every sector (Kay & King, 2020). Not only are these problems pervasive (and of central focus to this work), but many critical problems may further be considered '*super-wicked*' (Levin et al., 2012) problems, typified by more acute complexity and urgency. Such problems are marked by rapidly closing windows of opportunity, a lack of centralised authority to effect change, stakeholders who contribute to the problem being necessary to possible solutions, and cognitive biases like hyperbolic discounting that impair long-term planning.

Although organisations can in principle leverage greater resources than individuals to tackle these challenges, they also face challenges that offset, or even undermine these potential advantages. These challenges include:

- *Information*. Larger scale planning and decisions, regardless of time-horizons, often entail the incorporation of more sources and items of information (i.e., there are more "moving parts" to consider). Again, these add to the difficulty by increasing the number of potential parts to integrate into a coherent whole, each bringing uncertainties that interact and compound with others. This distributional challenge can also be compounded by potential miscommunication problems (e.g., the ambiguity of interpretation of the conclusions from other departments/teams, or even the same team at later time-points).

- *Incentives*. Not only may there be more affected parties for a given choice of action - each with potentially conflicting goals and incentives to consider, but even within an organisation there may be myriad perspectives and incentives to attempt to balance. These can all impact the difficulty of reasoning towards a decision.
- *Agreement processes*. Relatedly, the processes of collective reasoning and decision-making within an organisation may be dictated by any number of extraneous factors: hierarchy and authorisation processes, traditions, collective responsibility obligations, and more. These processes are not necessarily accuracy-aligned, and consequently can make arriving at a well-reasoned, accurate position difficult.
- *Time delays*. Finally, larger scale decisions may take longer to put into action, whether due to simple logistics, necessary oversight (e.g., budgetary, ethical) or other extraneous factors. Regardless of source, time delays add an increasing uncertainty in that the information basis of the original decision becomes more and more obsolete.

These factors all make RUU more difficult, and the increased number of considerations all come with aforementioned sources of uncertainty, which once again not only affect reasoning difficulty in isolation, but also in explosive combination. Of particular note is how these factors make *getting clear feedback* on the effects of previous decisions even more difficult - something we have noted as compromising the capacity to calibrate and improve reasoning outputs in section 1.2.3.

Finally, when we consider the effortful process of RUU, it is important to note that organisational decision-making often has more on the line, whether a government department advising national level policy, a multi-national business determining strategies that will affect thousands of jobs (and their competitive advantage / market share), or militaries determining effective strategies for hot or cold conflicts. On an organisational level, the pressure to get things right - or even just gain an *edge* over others - is a notable incentive for gaining any potential reasoning accuracy benefit. When we consider this in combination with the size of the reasoning challenges faced by such organisations (including the corresponding degrees of uncertainty and complexity, and the complicated and distributed processes and actions that may be involved), we should not be surprised by a) the persistence of potential mistakes, but also b) the demand for technical assistance.

## Chapter 2: Organisational Difficulties, Incentives, and Appetite for (AI) Assistance

### *2.1. Call for Assistance*

Organisations in both public and private sectors who have been facing these reasoning under uncertainty challenges have not been idle. In recent history there have been widespread calls for assistance, with applied efforts manifesting through various initiatives, grants, and technological ventures. For example, the US Intelligence Community recognised potential shortcomings in their provision of accurate intelligence post 9/11 and Iraq intelligence failures (see e.g., Robb, 2005), and launched a series of programs inviting outside assistance (e.g., ICD 203; US ODNI, 2015). The Intelligence Advanced Research Projects Activity (IARPA) launched the Aggregative Contingent Estimation (ACE; Matheny & Rieber, 2015), and the Crowdsourcing Evidence, Argumentation, Thinking and Evaluation (CREATE; IARPA, 2016) programs, the former seeking potential methods of aggregation across analysts/reasoners, and the latter seeking assistive tools.

Philanthropic organisations have also recognised potential shortcomings in institutional decision-making, echoing IARPAs funding initiative across the wider governance and policy space (see e.g., [philanthropic calls for improved institutional decision making](#)).

There has also been an increasing appetite within the private sector for augmented intelligence tools, particularly in fields where uncertainty is a constant companion. The rise of quantitative analysts—or "quants"—in finance is a prime example of this trend. Companies operating in volatile markets have turned to these specialists, armed with advanced mathematical and computational skills, to navigate through the fog of uncertainty.

It is worth noting that the human propensity to augment intelligence in the face of complex decision-making is not a recent phenomenon. Even in everyday scenarios, such as deciding on purchasing a house, individuals instinctively turn to tools like lists, spreadsheets, and calculators to augment their decision-making capabilities. These tools serve as extensions of our cognitive faculties, helping to structure information, compute numbers, and visually represent data in ways that enhance our ability to reason under uncertainty.

Within the domain of augmented intelligence, there is a broad spectrum of applications, ranging from strict tasks with clear remits—such as communication assistance, memory storage, organisation, and mathematical calculations—to more nebulous, complex, and multi-criteria decision and prediction tasks. The latter often relies on data-driven algorithms capable of leveraging vast datasets to make statistically inferred predictions (see e.g., Supervised Machine Learning). These have found applications in areas like insurance, stock trading, and actuarial sciences, where the abundance of data allows for the refinement of predictive models.

However, the landscape shifts when we venture into data-scarce, uncertainty-ridden problem spaces. In these domains, RUU becomes heavily reliant on human judgement. As a result, the augmented intelligence tools designed for such environments tend to be "hybrid" systems, requiring substantial input from expert human users. Despite these challenges, there have been some successful applications and recognized demand for such tools.

These methods can be classified in terms of Bayesian (e.g., Bayesian Networks, Pearl, 1988), and non-Bayesian approaches (e.g., Dempster-Shafer Theory; Shafer, 1976). Though both approaches have their proponents and detractors, their intent and core premises are sufficiently similar for our consideration: To reason under uncertainty effectively requires the decomposition/structuring of the sources of uncertainty in the problem; the quantified representation of that uncertainty, and the mathematical integration/combination of these components. Bayesian methods, for instance, have shown demonstrable value as normative benchmarks against which human reasoning processes can be measured and quantified (see e.g., Pettigrew, 2016).

Practical implementations of these hybrid systems can be observed in various sectors:

- In intelligence analysis, the CREATE program from IARPA has given rise to augmented intelligence tool projects like SWARM (van Gelder & de Rozario, 2017) and BARD (Nyberg et al., 2022) which each leverage combinatorial and computational methods to assist analysts in their reasoning tasks.
- In healthcare, tools like causal Bayesian network applications have been integrated into decision-making processes within the UK's National Health Service (see e.g., Babylon Health; see McLachlan et al., 2020).
- The realm of criminal justice has seen the adoption of tools like the Offender Assessment System (OASys) in the UK prison service, aimed at predicting recidivism (Moore, 2015).

Despite these strides, the integration of augmented intelligence for RUU in uncertain environments is not without its challenges. Often, hybrid systems, which balance automated reasoning with human judgement, emerge as the most effective solution. This can be attributed to the limitations of augmented intelligence - such as data limitations, or the need to have a human actor in the loop, ensuring accountability (for a review of hybrid systems, see Zheng et al ., 2017).

However, these systems are rarely plug-and-play; they require substantial training and integration of the human "components." This steep learning curve, coupled with the inherent complexity of the tasks at hand, means that the adoption of such tools comes with significant costs. For workers, it may demand a reconfiguration of skills and adaptation to new workflows. For organisations, it necessitates an investment of time and resources, weighed against the perceived benefits of enhanced efficiency, accuracy, and potential cost savings. Further, given the potential costs of errors in the decision-spaces to which these tools are being leveraged (e.g., intelligence errors, misdiagnosis, erroneous parole decisions), there is a pressure to ensure any analysis/reasoning provided by tools remains *explainable*, to ensure decisions are justifiable and errors detectable - something more broadly of concern to the AI space as a whole (Arrieta et al., 2020).

To understand organisation appetite for AI tools, along with an understanding of the historical calls for, and challenges facing, augmented intelligence solutions, we should next consider the more structural pressures and incentives on organisations.

### *2.2. A note on organisational incentives*

Moving from individuals to organisations introduces a plethora of incentives and priorities that guide actions and strategies, especially when grappling with decision-making about future actions (i.e., those that require RUU). Understanding these determinants is crucial not only from an analytical perspective but also when contemplating the introduction and adoption of technologies such as assistive AI tools. Priorities and incentives in organisational decision-making include:

- **Efficacy**: Achieving desired outcomes efficiently is fundamental. Any new solution or change to current decision-making must carry sufficient prospective gains to warrant the inconvenience and costs of that change to current processes. This incentive will vary with the organisation's perception of current efficacy levels (and other considerations below).

- **Cost**: Budgetary considerations always loom large. Reducing costs while maintaining or enhancing quality is a perennial objective. Investments in technology, particularly, are often scrutinised from a cost-benefit standpoint, and can include changes to operational costs (e.g., this software cost, vs current tools), and investment in labour (e.g., changes to necessary training / onboarding processes).
- **Smoothness**: The ability to maintain operational fluidity and minimise disruptions is vital, especially when implementing new technologies or processes. The more burdensome the prospective change, the lower the incentive to adopt it. Disruptions can include transitional periods of partial adoption (and therefore potential complications internally between existing and new processes), aforementioned training times and possible hardware/software installations.
- **Speed**: Across all organisations, speed in decision-making and forecasting is crucial. Whether in determining policy, military plans/responses, or business strategy changes, often these decisions require rapid assessment and response to changing conditions and finite windows of opportunity. Prospective changes to decision-making processes (including assistive tool adoption) face a fundamental barrier if their effective deployment is not sufficiently agile to accommodate this constraint.
- **Accountability**: Although the degree to which accountability is readily determinable varies from domain to domain, and decision to decision, broadly, organisations are interested in the determination of how or why a decision was made, and by whom. This may be for internal (e.g., promotional or disciplinary pressures) or external purposes (e.g., public enquiries, shareholders), but typically is closely tied to the internal and external criteria and rules that govern the decision-making process. This consideration is notable in the context of discussions of explainability within AI tools (Arrieta et al., 2020).
- **Reputation**: Organisations broadly must consider the potential consequences of decisions taken to their reputation. Whether this is the potential risk to future business, capacity to attract investors, pose a credible threat (e.g., militarily or regarding belief in enforcement), or prospective political/democratic consequences. Perceived incompetence or lack of trustworthiness from bad decision-making can have both immediate and long-term consequences to organisations.
- **Time-horizons**: Finally, it is important to consider that the time-horizons for organisational decision-making can vary substantially, and that these will interact with other incentives. For example, longer time-horizon actions may be discounted in importance relative to more immediate potential consequences (cf. hyperbolic discounting; see e.g., Frederick et al., 2002). Consequently, where a decision-process typically lies on this dimension will impact the organisation's appetite for innovation and change.

Taking a structural perspective, the introduction of any technology, especially a General-Purpose Technology (GPT) like AI, can interact complexly with these priorities. Historical patterns of GPT adoption reveal that incentives are intricately linked to embedded cultures, perceived threats or competition, and associated costs (Lipsey et al., 1998). Internal decision-making is often influenced by external forces and changing conditions. For example, when organisations sense an opportunity to gain a competitive edge, this can create pivotal moments for technology adoption. The innovation diffusion literature suggests that a kind of social contagion feedback loop emerges—a fear of missing out (FOMO) can intensify, attracting investments, talent, and more. This adoption dynamic is shaped by:

- **The balance between perceived advantage and risk**: Once an innovative technology is seen as offering more benefits than associated risks—and absent any major prohibitive factors like high costs or stringent regulations—there can be rapid, widespread adoption.
- **Heterogeneity in risk appetites and innovation capacity**: This diversity drives the diffusion process. Risk-tolerant firms might be early adopters, experimenting with new technologies ahead of their peers. In contrast, large, established firms, with more at stake, may be more cautious.
- **Competitive markets catalyse adoption**: In sectors with intense competition, the varied approaches to risk and innovation can accelerate the adoption of new technologies. Once the benefits of the technology are evident, or at least its value is proven, the larger incumbents usually follow suit.

This dynamic isn't confined to a single market or economy. With a GPT, adoption can ripple across sectors and borders, magnified by global economic linkages and the impacts of status and credibility. For instance, if prominent governments like those of the US or UK adopt an approach or technology, it can act as a catalyst for other countries to follow suit (see e.g., Behavioural Insights and Data Science departments; Sanders et al., 2018). Similarly, if a reputed company in one industry embraces a particular technology, it can signal to companies in other industries that the time is ripe for adoption. Within the realm of social dynamics, such entities act as hierarchical "influencers" (for a review on related innovation diffusion, see Peres et al., 2010).

In summary, the intricate interplay of organisational incentives, priorities, and the potential adoption of assistive technologies like AI tools is a testament to the complexity of decision-making in uncertain environments. In considering the prospective adoption of a new technology, one factor stands out as it pertains to AI tools: Easy to use and deploy tools, where costs are low or net positive (e.g., saving human labour), are generally more likely to satisfy organisational needs - particularly if efficacy is perceived (whether correctly or not) to be at least equivalent to current practices.

*2.3. The Appetite for LLM*

Given the incentives within organisations to maximise efficiency and solve for mixed-incentives that do not necessarily prioritise accuracy above all, it is perhaps unsurprising that augmented intelligence methods have had mixed success. Whilst assistive softwares have been integrated to manage other task domains (e.g., excel for logistics, email for communication), their uptake has typically been governed by incentive alignment: specifically, they have proven themselves cost savers (e.g., in employee hours).[2] Augmented intelligence softwares aimed at improving RUU functionality, in contrast, are typically training and use intensive, costly, and not always viable (see Giest, 2020). This is often because of the necessary restructuring of individual and group decision-making processes within organisations to see improvements to a previously nebulous process, often having previously developed and embedded organically (e.g., individual analyst methods, then circulated via prepared documents and discussed in a group meeting). This situation is perhaps unsurprising given the absence of formal RUU training and software norms across domains.[3]

---

[2] This is not to say assistive softwares have not been employed or ignored for extraneous reasons, such as cultural norms (e.g., for innovation, such as in the tech industry), or security concerns (e.g., development of in-house alternatives, such as HP internal internet).
[3] The exception to this are applied domains in which incentives and capabilities already align with the quantified augmented intelligence methods, such as actuarial work and risk-related industries like insurance - where inaccuracy can be quantified and dictates organisation survival. We argue that many organisations do not have

Consequently, to the naïve user, the advent and popularity of LLMs appears to be a potentially cost-effective, user friendly (and thus more incentive aligned) solution to RUU needs. Moreover, as things stand, the user friendliness and convincing nature of LLM outputs (see e.g., Palmer & Spirling, 2023) is only likely to increase this potential disconnect between naïve user implementation/trust, and underlying reasoning risks/errors. Such a temptation has, for instance, been found in the courtroom, with a lawyer caught citing fake cases mistakenly believed to be real due to their generation by ChatGPT (Sloan, 2023). This suggests an existing potential misuse rate may be higher than anticipated in the myriad domains without such close oversight/scrutiny.

This naïve user demand has a growing evidence base. Whilst the general appetite for LLMs like ChatGPT continue to grow (e.g., an increase of over [120% in ChatGPT use globally in the first quarter of 2023](#)), this has also affected high-power / important decision-making processes within industry and business ([80% of C-suite executives have exposure to AI tools, with 25% already using them in their work](#), Chui et al., 2023; [87% of Chinese workers welcoming AI tools to assist the key decision-making](#), BusinessWire, June 2023) including their deployment in forecasting and strategy tasks ([5% of C-level users already report using LLMs for forecasting](#), Chui et al., 2023). Moreover, this trend can be expected to increase, with business advocacy for prediction uses (see current [high profile "pitches"](#), Agrawal et al., 2023), and expectations of further implementation ([37% expect AI generated inputs to key decision-making](#), Laouchez & Misiaszek, 2023; and [35% to 65% of CEOs see reductions in strategy, supply chain, risk, and finance workforce/costs](#), Chui et al., 2023).

This current demand trend does not stop at the private sector. Governments and policy-makers have also shown an increased appetite for using these tools (e.g., [US Congress purchasing 40 ChatGPT licences to assist with policy-making](#), Krishan, 2023) and there is increasing advocacy and expectations for their use in governments across Latin America and the Caribbean ([OECD, 2023](#)), in economic policymaking in Asia-pacific governments ([APEC, 2022](#)), and [global advocacy for their directed use in "turbocharging" government forecasting](#) (Patel et al., 2021). Moreover, public sector focus on AI risks has been on data privacy, content legality, and generalised misuse (see these [US](#) and [UK](#) examples). It should lastly be noted that advocacy for the increased use of AI tools within the Intelligence Community have caused concerns regarding inappropriate use and over-reliance (see e.g., O'Brien, 2022; Treverton, 2023).

Having established the growing appetite for AI tool assistance in RUU problem spaces across organisations, noting its game-changing reduced barriers to adoption (e.g., low user effort and cost), we must now outline the true size of the challenges facing these tools, and from there determine potential (structural) risks.

---

the benefit of these strict accuracy incentives (which leads to the selection of employees with amenable skills), such as policy-making.

**Chapter 3: Reasoning Under Uncertainty, Wicked Problems, AI and NP-Hardness**

*3.1. What makes RUU hard?*

Fundamentally, there are two interrelated challenges that make RUU hard: combinatorial explosiveness[4], and compounding uncertainty. Both of these challenges scale at a non-linear rate with the "size" of the reasoning problem at hand. Whilst smaller scale, more constrained problems are consequently far easier to deal with (i.e., reasoning accuracy is more readily attainable), as problem dimensions grow in number, the difficulty ramps up exponentially. To elucidate:

Let us first consider a "simple" problem: should I take an umbrella for my walk in the park this afternoon? For this, I consider two relevant pieces of information: First, how often does it rain in November in London? Answer: Roughly about 1/3rd of the time. Second, the recent weather report stated that it will rain in London this afternoon. At this juncture, it is critical to note that my problem entails more uncertainties than simply my base rate of rain (33%) and my belief in the accuracy of forecast. For example, there is uncertainty regarding:

- Which areas of London is the forecast referring to?
- How does the accuracy of a forecast change as an event approaches?
- Is this particular source accurate?
- How is this accuracy distributed (e.g., conservative bias, random error)?
- Is my base rate appropriate at *this specific point in November*?
- Is this rain so bad I will not want to walk at all? Or so light I could wear a light coat instead?

Although *some* of these uncertainties I could reduce with further (empirical) enquiry with varying degrees of difficulty (e.g., checking if the source is reputable, gathering more forecasts)[5]. The main point here is that given a quick cost-benefit, I can be satisfied with integrating my two items of evidence together in a rough manner. Or put another way, I can be satisfied in not committing effort to resolving outstanding uncertainties much further: Rain is not unusual in November, and weather forecasts are generally accurate when predicting rain - to determine rain is more probable than not, and thus I should take my umbrella and not waste any more effort on the problem. The trade-off between cognitive and environmental effort and constraints and the value of the problems being addressed has long been established (e.g., heuristic rule application vs analysis, see e.g., Simon, 1955; Gigerenzer & Goldstein, 1996). Here we seek to emphasise that even seemingly simple RUU problems do involve many problem and decision dimensions (each with requisite uncertainties), but that in this case their low value justifies avoidance of further engagement with this difficulty. Example ways in which difficulty (and uncertainty) is excised include:

- There will have been uncertainties of which I was simply unaware.
- Some of the known uncertainties I elected to ***ignore***.

---

[4] This combinatorial explosiveness, as we will outline below, cuts across multiple decision-relevant dimensions, including the decision-space (possible actions/decisions), explanatory models, and sources of information. Work in Decision-Making Under Deep Uncertainty (DMDU), and work in Intelligence Analysis more generally has sought to address these dimensions via distinct approaches tailored to the particular issues entailed by each (for a summary of relevant approaches see e.g., Vincent et al., 2019). Here we take the more generalist perspective that RUU problems can involve *any and all such dimensions* when considering task difficulty as it pertains to both human and AI performance.

[5] In fact, for this specific question, many of these uncertainties could be addressed via detailed rainfall pattern data for the forecast time period (assuming the person is aware and has access to such information). Our point here is that resolving/reducing uncertainties requires further time and effort, with varying degrees of success.

- I did so based on a very simple cost-benefit trade off - is my expected gain (the benefit of the prospective increase in accuracy) worth the time and effort.
- My choice was simple, take or leave the umbrella, and the outcomes were clear (get rained on, or stay dry).
- My integration of the facts was fine to leave rough - take or leave the umbrella is binary, and given being without an umbrella when you need one is worse than having one and not needing it, my two items of information were sufficient to raise my probability of rain enough (e.g., over 50%).
- I have had to make this decision repeatedly in my life.
- I can expect the broad underlying processes to remain the same (e.g., my umbrella will continue to be effective, weather will continue to be forecastable by current models/meteorologists).

These are just a selection of ways in which my reasoning process either excised, or avoided dimensions of uncertainty. Some of which directly assisted my accuracy (e.g., stable processes, repeated experiences) whilst others carried no meaningful *detriment* to my accuracy (e.g., cutting off further information search).

In contrast, let us consider a reasoning problem that is slightly "larger": How should I (a policy-maker) best protect my country against the risks of AI? For this, there are substantially more components to the problem. For example, the current status of AI tool development, the rates of development to date, possible hardware, energy, and economic constraints, current regulatory approaches, and literature on current and possible risks associated with development and use, existing potential policy levers, are all relevant to the question at hand (and are by no means an exhaustive list of considerations). Two immediate challenges arise:

First, how do I put these components together (i.e., model them) to formulate a conclusion? When we had a "2 body" problem, our possible ways of combining the items of information was made simple (our one piece of evidence updates our original/background belief). The moment we have more than two items under consideration, our number of ways of combining them grows exponentially (for relevant historical discussion of these difficulties from the field of Intelligence Analysis, see e.g., Heuer, 1999). For example, whilst the relationship between hardware capability and AI tool development may initially appear straightforward (e.g., hardware increases enable AI tool development increases), how background economic conditions and regulatory conditions interact with those two components (and their potential relationship) is far less clear. Whilst causal schemas (i.e., external knowledge we bring to the problem that guides our understanding of likely causal relationships) can help us in constructing *some* elements of the problem (e.g., we know hardware cannot run without energy, development cannot exist without hardware), the number of *potential* combinations still poses a substantial challenge[6]. Moreover, and as highlighted in figure 3.1.1, the number of variables (information items) and their combination are just two of *many dimensions* that dictate the number of possible worlds to consider.

---

[6] The challenges posed by structural uncertainty in particular, such as causal relationship complexity, and uncertainty quantification difficulties have long been recognised, with attempts to address it including expert elicitation (see e.g., Cooke, 1991), sensitivity analysis (Saltelli et al., 2008), causal discovery algorithms (Spirtes et al., 2000), and hybrid approaches (Oreskes et al., 1994), among others.

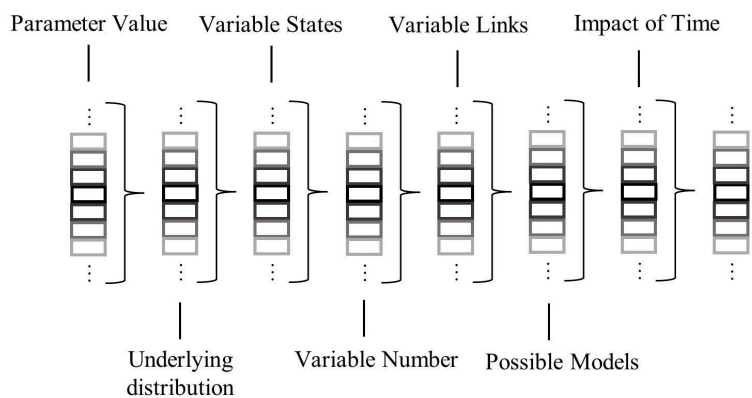

*Fig 3.1.1. Combinatorial sets of reasoning dimensions. When considering possible <u>actions</u>, additional layers of dimensions are also required, including possible form, combination, and timing of actions.*

To illustrate the potential computational explosiveness further, let us consider the fact that each of those items of evidence may have more than two possible "states" it can occupy. In our simple problem, each item could be stripped down to 2 states: rain or no rain. Thus, the total possible world corresponds to the number of possible input values, to the power of the total number of inputs required. If we assume determinism, then this is $2^2 = 4$ possible worlds. Of course, when we begin to consider uncertainty, by incorporating possible probability values for each input value (and the possible distributions these are drawn from), then even this number begins to grow exponentially.

In our more complex version, we have many more variables that may have myriad possible states (e.g., levels of energy output; possible regulatory policy positions), and critically *all of these states may interact with all of the potential states of any other variable*[7] - including in conjunction, which necessitates multiplication, rather than addition of sets of inputs. For example, even if we arbitrarily constrained hardware to 10 possible states, and regulatory policy positions to 10 possible policies, and we want to know how these may affect the 10 possible rates of AI development, then just this small part of the problem (which we take pains to note we have made substantially simpler by excising all other interactions with the remainder of the problem), then we must now consider $10^{((10 \times 10)+10+10)} = 10^{120}$ possible worlds!

We can then quickly see how considerations of which probability value to assign to a state or conditional relationship, and from which underlying distribution, multiple possible models, possible underlying distributions that govern the probability of states within a variable, or relationships between states in variables, the way any of these dimensions may change *over time*, will *all contribute* in multiplying this rapidly growing number of possible worlds to consider. This combinatorial explosion has led to the conclusion that the amount of necessary compute can quickly approach an unfeasibly large number (though we note approaches such as Monte Carlo Rollout have made progress in navigating dimensions of this challenge; see e.g., Silver & Veness, 2010).

Here, however, is the second part, where the RUU problem gets even trickier. In many instances we do not (and *cannot*) know the possible values of certain dimensions. For instance, we do not know which variables that may or may not be relevant to the problem remain unknown (i.e., unknown unknowns). Revisiting figure 3.1.1., we can think of this as not knowing the full "stack" height of the "variable number" column. Critically, because dimensions are multiplicative, this deep uncertainty *spreads* across the set of possible worlds. If, for example, we were in fact missing 50% of the key variables (again, we emphasise here that part of the problem is we fundamentally cannot even determine this percentage), ceteris paribus we are missing 50% of the possible world upon which to perform our conditionalization calculations. Critically, because these dimensions are all *interactive*, we cannot abstract away this influence (e.g., by arbitrarily stating now our conclusions are half as likely as they were before), because there are no guarantees that the appropriate underlying distribution is symmetric (i.e., there may be greater or less than 50% of the final conditionalized probability distribution in the unknown half of the possible worlds - *even when we know we are missing exactly half, which we also can't know*).

---

[7] We note that although decision options do not often impact remaining probabilities (i.e., uncertainty states across other variable/state combinations), a) they are capable of doing so (and we wish to argue such a possibility as of valid concern), and b) the multiplicative impact across remaining variable/state interactions remains sufficient to warrant the combinatorial explosiveness described here.

These types of unknowns / uncertainties have been termed as deep (Lempert et al., 2003), severe (Ben-Haim, 2006), Knightian (Knight, 1921), and radical (Keynes, 1921). Their core principle is the same: they are a class of uncertainty that is not reducible to a quantified state (e.g., a probability). For example, whilst we may be able to infer aggregate or "typical" futures that are produced by generalisable rules (quantified or stochastic uncertainty), we cannot infer a *specific* future event that is the product of that is the product of a complex system[8] (particularly if those events are in any degree dependent on human behaviours and beliefs). So whilst we may be able to estimate, for instance, whether or not an AGI might be developed in future, we cannot predict the specific event itself (i.e., who, where, how, when). This would entail knowing about *every* surprising, unexpected, or intervening factor that occurs between now and that event, *and their interactions* with the reasoners model of the system - even if these events are unprecedented.

In more formal terms, this deep uncertainty has been defined as:

"the condition in which analysts do not know or the parties to a decision cannot agree upon (1) the appropriate models to describe interactions among a system's variables, (2) the probability distributions to represent uncertainty about key parameters in the models, and/or (3) how to value the desirability of alternative outcomes."

*(Lempert et al., 2003)*

Returning to the issue of RUU in the face of *wicked* problems, whose characteristics include uniqueness, multiple possible explanations, and no clear formulation (Rittel & Weber, 1973), we can see how pervasive deep uncertainty is in the problem by definition (i.e., there is simply an absence of information to even begin reducing the uncertainty around, akin to trying to make a prediction based on one data-point). The obvious question that then arises is if this uncertainty is irresolvable, what can be done about it?[9]

---

[8] In fact, DMDU proponents would caution against attempting such an inference. Whilst here we seek to illustrate the principles of deep uncertainty resulting in fundamentally unknowable predictions, those who have long focussed on actionable solutions to deep uncertainty difficulties have argued that these questions are not necessary for reasonable decisions. Instead, we should pay attention to how far apart these futures are in the decision-space (i.e., a focus on alternative metrics, outside of the precision focus here, such as robustness; see e.g., Lempert et al., 2003).

[9] It is again worth noting, that DMDU approaches, such as Robust Decision-Making (Lempert et al., 2003) instead focus on robustness among possible options, rather than trying to seek out a clear optimum. However, here we are trying to illustrate the breadth of the challenge in principle, before later diving into possible mitigations, circumventions, and solutions (see Chapter 6).

In human reasoning, this form of uncertainty is not typically a conscious consideration, but rather is effectively ignored relative to a reasoned action at hand. This may be via the implementation of basic causal schemas (Cheng & Holyoak, 1985), heuristics and shortcuts (Tversky & Kahneman, 1974), or more broadly as a satisfactory margin of error, given cognitive and environmental constraints (e.g., Gigerenzer, 2002; Oaksford & Chater, 1998), though it should be noted this remains underexplored. Critically, this error may be efficient in the everyday reasoning problem spaces, but is not fitted to larger scale or *wicked* problems.

On a larger scale, scientific and economic thinking has attempted to excise the consideration of deep uncertainties within models and frameworks (e.g., probabilistic reasoning, modern economic theory; e.g., Von Neumann & Morgenstein, 1944), despite exceptions (Knight, 1921). However, those who have focussed their attention on this issue for decision-makers (and notably those faced with long-term policy and strategy decisions) have forwarded frameworks that acknowledge the disruptive impact of deep uncertainty. For example, 'adaptive foresight' is an approach to policy based on a reactive, adaptive approach to ever-changing conditions (Eriksson & Weber, 2008). Instead of attempting "impossible" prediction, the 'constraint' is relaxed to be about dynamic response to the constantly changing information environment.

In sum, the RUU combinatorial explosiveness issue, and associated deep uncertainty challenge become increasingly more problematic as we consider larger, less constrained problems, whether it is longer time-frames, the inclusion of complex/interdependent components (e.g., human behaviour based on other human behaviour), or wider potential decision-spaces (among many other dimensions). Whilst the non-linearly increasing computational challenge that accompanies the growth of these dimensions already poses a substantive barrier to accurate predictions, it is the creeping, non-linear growth of deep uncertainties that stand to fundamentally compromise how *knowable* a future is - even with overcoming computational difficulties. This degree of irresolvable error is an open question in the domain of human reasoners (made challenging by those same problem features), and raises the issue of whether AI tools can meaningfully improve upon this.

### *3.2. The RUU challenge for current AI Tools*

Having laid out the principle difficulties associated with RUU, and in particular relative to the larger problem and decision-spaces of interest, we now turn to the question of current AI tool capabilities in this regard.

First, while the current evidence base is growing rapidly, at time of writing evidence of RUU capabilities are limited. Whilst various LLMs have demonstrated notable strengths in logical (Creswell et al., 2022), mathematical (Imani et al., 2023), and semantic reasoning (Tang et al., 2023), among others, these demonstrations are all within rule-based domains that are bereft of the uncertainties we have been discussing. Moving towards RUU related capabilities, such as planning (Valmeekam et al., 2023) and causal reasoning (Jin et al., 2023), evidence suggests some limited capacity to engage in these processes, *under highly constrained conditions*. This has been further confirmed by direct investigation of forecasting ability, wherein GPT-4 underperformed in forecasting accuracy relative to human controls on questions that by definition apply substantial constraints over the problem space (Schoenegger & Park, 2023). Outside of highly constrained reasoning problems, it appears current models - even with the assistance of protocols expected to enhance reasoning fidelity (e.g., Chain of Thought prompting; Nye et al., 2021) not only behave inaccurately and inconsistently in the face of these challenges, but in fact larger models *perform worse* (Lanham et al., 2023), raising questions over compute margins being the solution to RUU performance (a concern in line with arguments laid out in Section 3.1).

Second, whilst direct evidence is not encouraging, the question then arises of theoretical capability in principle. In this regard, we must first acknowledge the unpleasant computational truth that RUU in unconstrained problems is in nondeterministic polynomial time (though laid out in basic form in Section 3.1., for full treatment see, Dagum & Luby, 1991; but also Cooper, 1990; and Geist, 2020), but this is not just in determining exact computation, but even in approximate computation. It is worth reiterating here the core combinatorial explosion issue: for each dimension that increases the number of required inputs (which should be noted as a generally multiplicative calculation; e.g., more variables, possible connections between variables, temporal changes and process changes) the resulting sum is an exponent upon the number of possible values each of those inputs could take.

$$Number\ of\ Possible\ Worlds\ =\ \prod_{i=1}^{n} k_i$$

Where *n* is the number of variables, and each variable *i*, can be in one of *k* states, such that $k_i$ can be different for each variable. Thus the number of possible worlds is a product of each variable state combination. Of course, this assumes variables are independent of one another - which is a simplifying assumption that excises further complexity from the real world.

Consequently, this combinatorial explosiveness results in a conditionalization across sets of possible worlds in orders of magnitude greater than the number of atoms in the known universe[10]. If we take our ballpark estimate of $10^{120}$ possible worlds from Section 3.1 (note: an estimate based on only 3 variables with 10 possible states each - whilst other real world problems can easily reach 10 times the number of variables), and we assume one world only requires one computation, and we assume a generous dedicated FLOP of 1 exaflop ($10^{18}$ operations per second), then we are left with the following compute time:

$$Total\ Compute\ Time\ (seconds)\ =\ 10^{120}\ /\ 10^{18}\ =\ 10^{102}$$

Putting this into years[11], this translates to a time to compute of $3.17 \times 10^{94}$ years, with a correspondingly astronomical energy requirement and accompanying cost. Of course, the moderating factors here are a) what shortcuts can be taken to reduce computational requirements, such that b) accuracy (or approximate accuracy) remains sufficient for reasonable improvements over human equivalents?

Although our ballpark number of possible worlds from the 3 variable RUU problem above is already *notably constrained*, supplementary methods and features could help in reducing this number, such as applying functional mapping over sets of values for given variables, and/or applying degrees of selective ignorance to subsets of possible worlds. However, even with these sorts of techniques, that intend to relax the **exact compute** requirement in favour of **approximation**, *NP-hardness* still results (see e.g., Dagum & Luby, 1993). An open question is to what degree this combinatorial explosiveness can be addressed by quantum computing (something we will touch on later in Section 3.3), though even substantive orders of magnitude in potential operations per second (e.g., $10^{30}$) still leaves us in the realm of infeasible compute times.

---

[10] For context, the universe is ~$5 \times 10^{17}$ seconds old, and contains ~$10^{80}$ atoms.
[11] There are approximately $3.154 \times 10^7$ seconds in a year.

The question of whether accuracy of current AI tools *can* surpass human accuracy is an empirical one, and something that has yet to be proven in the affirmative (see e.g., Schoenegger & Park, 2023). Ultimately, such a comparison also rests on how well AI tools (vs human equivalents) deal with the second RUU problem.

Potential performance is also subject to the second, more fundamental barrier of deep uncertainty influences. Ultimately, regardless of computational capability of current (or future) systems, a ceiling on potential RUU accuracy is a function of the deep uncertainty in the particular problem space (which in turn is also relative to the question at hand). Put another way, even if one *could* perform a theoretical compute over all possible worlds, some unknown percentage of those inputs stubbornly remain unknown, and thus some unknown degree of inaccuracy remains. Categorically, we can characterise this as an AI failure of *practical impossibility* (Raji et al., 2022), such that there is insufficient observable data to construct an appropriate model (Jacobs, 2021; Jacobs & Wallach, 2021).

At present, evidence to date suggests current AI tools do not yet have the computational capability to brute force (or side-step) the necessary compute either to a) notably surpass human performance, or b) close upon the deep uncertainty ceiling. Though it should be underlined that this assessment is in part based on an absence of current capability awareness – which is an investigative imperative we argue for throughout this work.

### *3.3. The RUU challenge during transitional periods*

As we move from consideration of current systems towards more advanced, AGI-like systems, we must first consider the possible transitional periods of AI tool development. We first note in this regard that prospective timeline durations are less critical to our concerns than the form (or pathway) development takes. Consequently, timeline length (whether within the next decade, e.g., [Metacalculus forecast](), or substantially longer) is less relevant than pathway uncertainty (Gruetzemacher & Paradice, 2019), such as the high degrees of uncertainty not only on prospective feature developments, but on potential "caps" or limits on development, including hardware, resource, and energy expenditure (Kurshan, 2023; or for a more pessimistic outlook regarding the extrapolated reliance on unsustainable computing power gains, see Thompson et al., 2020). As we turn to considerations of AI development in relation to tackling the computational and uncertainty difficulties associated with RUU, we must again note that this is under a cloud of considerable uncertainty itself, and is therefore speculative at best. Regardless, a few general principles are worth noting:

First, in dealing with the computational challenge, it remains possible that sufficient hardware developments (and accompanying model sizes) can move towards outperformance of human reasoners, though this may be subject to the manner of development, and functional limits. For example, one way of describing this development challenge is that the problem task in which we seek improvements in performance (RUU performance on wicked problems) is one bereft of not only clear reinforcement learning signals, but also ready demonstration data from human performance (known as a 'Type 3', inelastic problem - see e.g., [Marblestone, January 2023]()). This is something to which the wider awareness amongst developers of the RUU challenge (part of the necessary solutions described in Chapter 6) seeks to remediate and accelerate. However, true advances in necessary compute times to start approaching deep uncertainty-imposed ceilings may be more feasible (and rapid) with advances in quantum computing, given the parallelizing capabilities can better address the polynomial compute times[12]. It is further worth noting in relation to the compute challenge, that the aforementioned potential hardware and energy caps/limitations in AI development (Kurshan, 2023; Thompson et al., 2020) are likely to impact progress in the most compute-intensive areas of function, of which RUU can be argued is one.

Second, and echoing a now familiar refrain, the deep uncertainty challenge does not have a clear theoretical solution. This being said, it remains an open question to what degree features may be developed (and solutions/uses refined) which allow for some degree of mitigation of the impact of this imposed ceiling.[13] This is of particular note given limits on available data to address problems via current methods (e.g., LLMs) - without sufficiently (and arguably impossibly) expanded available datasets, current development pathways will remain hamstrung for RUU performance in real world problems. We note that data limits are already a noted problem in AI development (e.g., Xue et al., 2023), with techniques including code incorporation and data filtering leveraged to attempt to reduce data-constraining limits to compute scaling laws (Muennighoff et al., 2023).

Regardless, whilst exact and approximate compute are not clearly solved by currently known trajectories and development timelines, it remains a more open possibility that future development towards AGI-like systems can reach a threshold of surpassing human performance. The exact timing and nature of this point is not only highly uncertain, but also subject to the application (or ignorance) of solutions to this issue, via directed evaluative and development attention.

### 3.4. The RUU challenge and AGI-like systems

Lastly, we turn to the advent of AGI-like systems themselves, and the question of whether such systems will be able to surmount the challenges raised here. Within this context, it is important to note the distinction between theoretical, ASI systems, and practically realisable, prospective AGI-like systems. In this regard, whilst ASI (e.g., AIXI; Hutter, 2007) are based on a theoretical notion of *infinite compute*, how one defines the "end-state" or threshold of an AGI-like system remains an unresolved question (cf. "Highly autonomous systems that outperform humans at most economically valuable work", OpenAI, 2018).

For an ASI, the current prospects are more straightforward: infinite compute should be able to handle the unfeasibly large compute implied by the combinatorial explosiveness problem of RUU. However, unknowable, irresolvable uncertainty as a challenge remains arguably untouched (and thus remains as an accuracy ceiling), and keeps *solvability* of RUU out of reach even theoretically (for an exposition of predictability limits for ASI, see e.g., Heninger & Johnson, 2023). However, as we have noted throughout, whilst deep uncertainty prohibits optimal accuracy, better-than-human performance remains more readily achievable, and within the criteria to meet AGI classification.

---

[12] As noted in Section 3.2, even theoretical orders of magnitude improvements can still leave the combinatorial explosiveness problem unresolved.
[13] As we will outline in the solutions chapter, *directed development focus* entailing the integration of DMDU insights may yield productive gains in AI tool efficacy.

For an AGI-like system, it remains plausible that whilst meeting (the albeit unclear) threshold for definition as an AGI (e.g., generally better than human performance), this does not guarantee supreme RUU capabilities. Ultimately, it remains possible that where performance is based on superior compute, an AGI-like system should outstrip human performance, but where deep uncertainty ceilings have greater impact (i.e., the ceilings are lower), this margin of improvement will be reduced. Put another way, like ASI, an AGI-like system will still be subject to the ***NP-Hardness*** issue invoked by deep uncertainty (see e.g., Lempert et al., 2003), regardless of compute, and thus will be subject to a margin of error that is at minimum a function of the degree to which deep uncertainty is inherent to the problem/decision space.

### *3.5. Summary: The RUU challenge and AI Timelines*

So where does this leave us with respect to an overall outlook on RUU capabilities in AI tools? And how does this relate to the overall problem at hand?

First, we have sought to underline just how *hard* RUU is as a capability, and though noting its difficulty is problem-dependent, the degree of combinatorial explosiveness and deep uncertainty influences set a high NP-hardness bar to pass for not just exact but even approximate solutions.

Second, we note that optimal or near-optimal performance is not necessary to outperform human reasoners on these tasks, and given the larger problem at hand pertains to relative advantage of AI tool use in this space, comparison to human performance is a key indicator.

Third, and relating to both these points, is the issue of evaluative unknowns. Specifically, we do not know enough about current and prospective AI tool performance on unconstrained (and even constrained) RUU problems relative to human performance. This is in part due to 1) the novelty of the evaluation field in AI tools and the amount of evaluators with both the necessary skill-set and model access (for a relevant summary, see Gruetzemacher et al., 2023), and 2) the inherent difficulties of assessing RUU, given the aforementioned challenges, also make clear feedback and faithful process assessments hard to come by. Critically, the evidence we *do have* suggests an assumption of capability would be misplaced.

Fourth, current evidence (limited as it may be) suggests inferior RUU performance - something that is yet to be comprehensively understood by those seeking AI tool application in this arena (see Section 2.3). Consequently, this is not only a prime driver of the full RUU Trap described below, but also a meaningful point of intervention.

Finally, development trajectories and AGI-like systems are by no means guaranteed to absolve us of the RUU problem. Whilst gains in computing capacity may assist in improving accuracy in relation to combinatorial explosiveness - and potentially provide a consistent advantage over human reasoners - the degree to which this will hold is dependent not only on the (guided) development of AI tools, but also the deep uncertainty intrinsic to the problems/decisions to which RUU processes are being applied.

Critically, an unknown/unchecked AI tool RUU error rate is one important part of a more substantial problem we will lay out in Chapter 4. The uncertainty surrounding (and consequent risk of) RUU errors we have described here, when considered within a framework of isolated independent risks, appears *potentially* costly (and worthy of correction), as a fixed misuse risk. However, when we place this misuse within the wider, structural risk context, we begin to see how interactions can elevate this risk to current, societal-scale levels.

## Chapter 4: The Reasoning Under Uncertainty Trap

### *4.1. Structural Risks and Compounding Effects*

When we consider the risks associated with AI development and use, those that stem from complex interrelations and chains of events within the surrounding *system* - structural risks are often overlooked (Weidinger et al., 2023; Zwetsloot & Dafoe, 2019). For example, whilst there may be focus on societal scale risks stemming from overreliance on an AGI-like system (and its potential misalignment), there is less focus on the chain of events - or *dynamics* - that have led the system to that point of dependence. In essence, structural risks stem from the interactions between components of the surrounding system, such as the rules or structures within a system that govern how actors within those systems are incentivised, and therefore interact with one another. Critically, when we take this perspective, we not only take into account the ripple effects, cascades, and other repercussions from an AI onto the broader system, but also how the structure of the system may shape subsequent AI development and actions (Zwetsloot & Dafoe, 2019). We note that this perspective gels with recent calls regarding the importance of "socio-technical" (human and system impacts) AI risks (see e.g., Lazar & Nelson, 2023; Weidinger et al., 2023).

The system dynamics that comprise structural risks lend gel with pre-existing complex systems thinking, where the interactions within a complex system have been investigated in terms of the variability, connectedness, interdependence, and adaptiveness among populations can produce emergent behaviours not readily deductible from base conditions (for an overview see Goldstein, 2011). The manner of these system behaviours have been classified into various phenomena, including feedback loops (positive / accelerative, and negative / stabilising; Forrester, 1961), buffers, time-lags, self-organisation, system rules, and goals (among others; see e.g., Meadows, 2008). Whilst simulation methods are favoured for thorough quantitative analysis of these systems (see Miller & Page, 2008), the complex systems approach lends itself as a theoretical lens through which to view structural risks within AI safety. Here we apply this methodology to understand the potential compounding feedback loops that result from the interactions between properties of the problems needing solutions, the human users attempting to find those solutions, and the tools they turn to in this need - all of which is embedded in the broader system context (e.g., mixed incentives). The interlinked compounds described here are argued to exacerbate the RUU risk stemming from AI tool misuse laid out above, doing so in a non-linear fashion. We note that whilst *some* of these compounds may also affect human RUU in a BAU scenario (something we take pains to note in comparison as we summarise relative impact), it is in AI tool development and use where we stand to make a difference - for good or ill.

### *4.2. The RUUT*

Over the preceding chapters we have established that a) there are many high impact, *wicked* problems that require RUU to address, with which human decision-makers currently wrestle, b) that there is a growing demand among prospective (high power) decision-makers to leverage LLMs for this purpose, and c) that LLMs are not known to be currently suitable for those RUU purposes, and in fact the deep uncertainties inherent to the *wicked* problems means even AGI-like systems may struggle to provide accurate assistance. Given this, we describe here the reasoning under uncertainty trap (RUUT) that stems from this predicament. RUUT relates to the umbrella risk category of "organizational risks" (see e.g., Hendrycks et al., 2023), but extends this to capture the compounding, structural risks (Zwetsloot & Dafoe, 2019) that amplify the potential costs in a "trap" of system feedback loops.

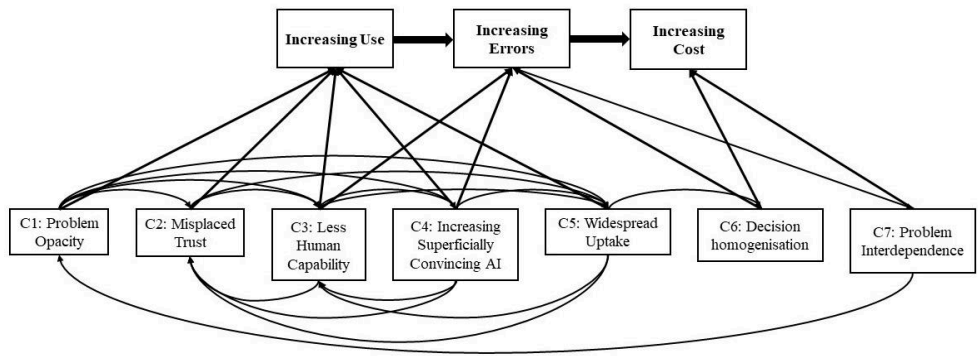

*Fig. 4.2.1. Map of Reasoning Under Uncertainty Trap (RUUT). From left to right (bottom row) are compounding factors (C1-C7), with arching arrows "forward" (top of boxes) representing feed-forward mechanisms, and arching arrows "backwards" (bottom of boxes) representing feed-back mechanisms. The central misuse chain (use to error to cost) is represented above, with direct impacts from compounds represented by straight arrows.*

Figure 4.2.1 illustrates the overarching RUUT set of compounds (bottom row) and their impact on the central costly use error problem (top row). Each of these compounds and their impacts with the rest of the system are highlighted and discussed in turn.

### 4.2.1. Problem Opacity (C1)

The forecasting problem has *deep uncertainty*.

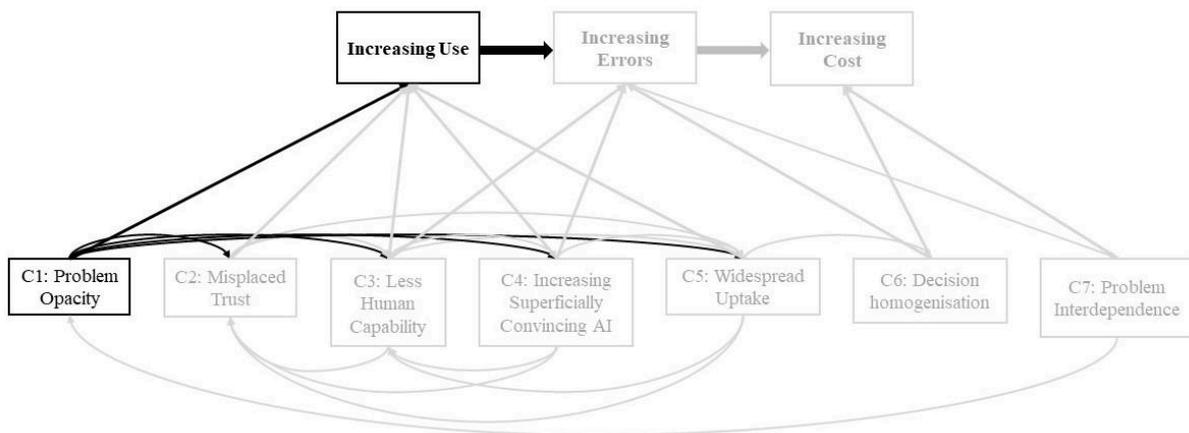

*Fig. C1. Map of Reasoning Under Uncertainty Trap (RUUT), highlighting Problem Opacity compound (C1). Feed-forward effect to further compounds highlighted in arrows arching forward from top of C1 box, with direct impacts on central problem highlighted with straight arrows.*

The core compounding factor is the issue of *Problem Opacity* (Fig. C1).

> ***Example***: *Government Organisation A is attempting to forecast future states of geo-politics to provide policy recommendations. The number of unknowns (other actors, shocks, unreliable/missing data, novel events, etc.) make accurate forecasting hard, and bereft of clear feedback. Analysts within Organisation A are tempted to turn to LLMs for assistance in the face of this challenge.*

Put simply, because of the deep uncertainty inherent to the wicked problems being faced (and more generally within the compounding uncertainties of RUU predictions), and notably the absence of clear feedback, erroneous decisions (absent effective and corrective RUU assistance) will not only be likely, but will be difficult to identify and correct. This difficulty is a direct contributing factor to increased levels of (mis)use as it is one of the incentives for requiring AI assistance in the first place (i.e., the challenging nature of the problems being faced drives human users to these tools). However, this factor additionally feeds-forward to the next four compounding factors – all of whom additionally impact (mis)use and error rates.

### 4.2.2. Misplaced Trust (C2)

Decision-makers start to trust AI tool forecasts, given lack of correction.

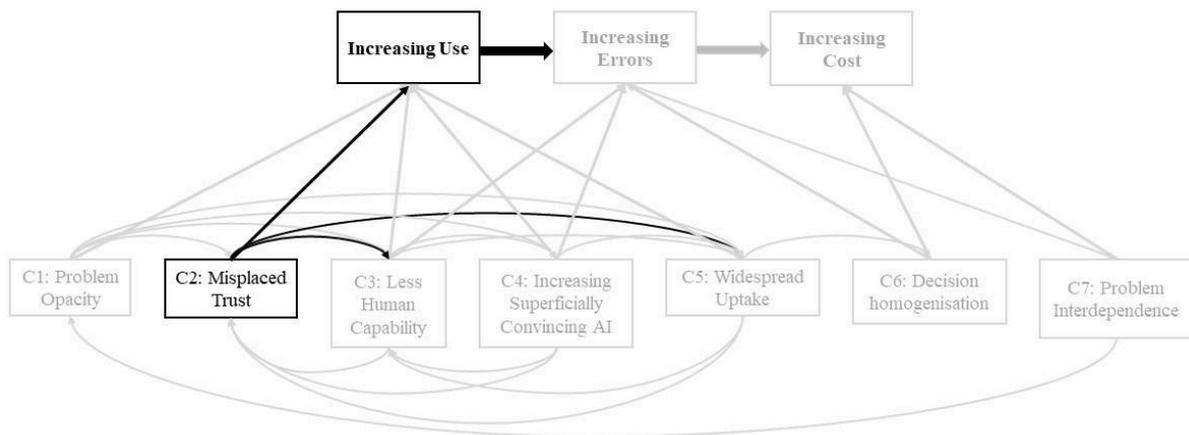

*Fig. C2. Map of Reasoning Under Uncertainty Trap (RUUT), highlighting Misplaced Trust compound (C2). Feed-forward effect to further compounds highlighted in arrows arching forward from top of C2 box, with direct impacts on central problem highlighted with straight arrows.*

As the problem opacity restricts learning opportunities, and AI tool use continues unabated, organisations, institutions, and individuals acquire a misplaced trust in the value and efficacy of such methods (for a related discussion on trust and AI alignment, see Liu et al., 2023). Conceptually, this relates to the AI failure concern of *falsified or overstated capabilities* (Raji et a., 2021), though critically the unearned trust is not necessarily a product of explicit developer claims, but rather from limited user perceptions. This misplaced trust in turn feeds back into further use of such tools, given their *apparent* efficacy – which proportionately increases the degree of erroneous decisions made (given this use is functional overextension). This in turn feeds forward into additional compounding factors across human user, AI tool development, and system dimensions.

It is also worth noting at this juncture, that trends in the automation of science (e.g., King et al., 2009) may also feed into this compound. Specifically, and depending on the nature of RUU requirements and oversight within science domains, automated science may in fact generate an evidence base that itself may be compromised, and yet incorporated into AI tool RUU processes. As a result, misplaced trust may in fact be more substantial when reasoned solutions rely on this material - which itself may become an increasingly likely prospect (see Chapter 5 for further details).

> ***Example***: *Organisation A has a few policy-makers that have been getting forecast material from LLMs that seems plausible. Combined with the lack of clear corrective feedback (C1), a misplaced trust builds in use of LLMs, with internal advocacy increasing use within Organisation A (C5).*

### 4.2.3. Less Human (RUU) Capability (C3)

Human oversight decreases as cost/effort seems less justified.

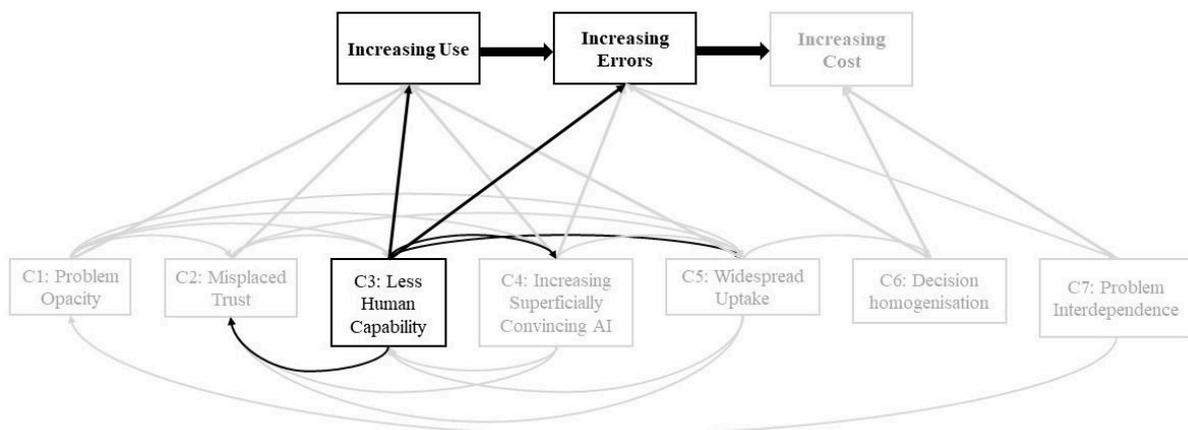

*Fig. C3. Map of Reasoning Under Uncertainty Trap (RUUT), highlighting Less Human Capability compound (C3). Feed-forward effect to further compounds highlighted in arrows arching forward from top of C3 box, and feed-back effects arching backwards from bottom of C3 box, with direct impacts on central problem highlighted with straight arrows.*

As AI tool use increases for RUU processes, absent intervention, the human RUU skillset – whether assisted by collective or (effortful) augmented intelligence frameworks – used to generate and *evaluate* RUU outputs (i.e., decision-making reasoning inputs) begins to wither. It does so as the apparent need for human oversight/involvement dissipates, and seemingly sufficiently high-quality outputs can be generated with relative ease. As the apparent need for human RUU capabilities in these problem spaces (i.e., organisations, institutions, businesses, etc.) lessen (see above section regarding cost reduction intentions), then 1) more dependence on AI tools is fostered to deal with said problems, and 2) investment in independent / non-AI oversight declines. These two aspects not only increase the amount of (mis)use, but also directly increase the risk of error from that use (see related organisational risk; Hendrycks et al., 2023). Although this factor will increase with problem opacity and misplaced trust (C1 and C2), there is a dangerous feedback cycle with subsequent compounding factors involving the route AI development takes, and the breadth of use (C4 and C5), where lessening human

RUU capability increase the dangers of these latter compounding factors, which in turn increase the rate of human RUU capability decline.

> ***Example***: As analysts in Organisation A trust LLMs for their forecasting tasks (C2), there is less of an apparent need for the time/effort/cost of continuing the more labour intensive (human) version, so costs are reduced and apparent efficiency is increased. However, the organisation as a whole becomes even more vulnerable to forecasting errors as oversight weakens — impacting not only necessitated increased use, but also the potential size of errors, and in turn increasing misplaced trust and convincing others within Organisation A to also use LLMs (C5).

### 4.2.4. Increasingly (superficially) Convincing AI (C4)

RUU-blind development leads to cheaper, more plausible, but no more accurate forecasting assistance.

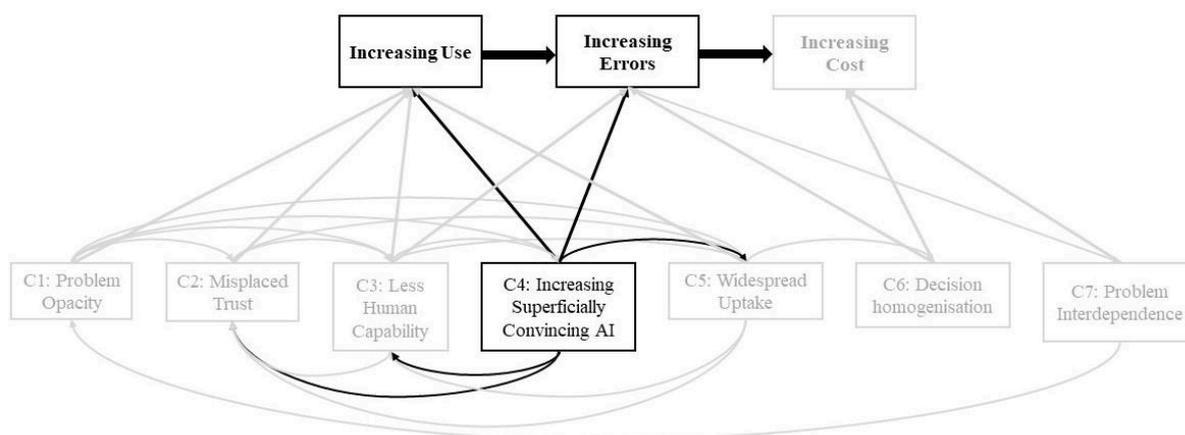

*Fig. C4. Map of Reasoning Under Uncertainty Trap (RUUT), highlighting Increasing Superficially Convincing AI compound (C4). Feed-forward effect to further compounds highlighted in arrows arching forward from top of C4 box, and feed-back effects arching backwards from bottom of C4 box, with direct impacts on central problem highlighted with straight arrows.*

As AI development continues, the outputs it generates seem increasingly convincing (see e.g., Palmer & Spirling, 2023). However, whether these outputs, in an RUU case, match up to an increase in *accuracy* is by no means guaranteed. In fact, absent intervention solutions (described below), the opaquer / more *wicked* the problem space (C1), the more users a) blanketly trust outputs (C2) and lack sufficient independent oversight capabilities (C3), the more likely AI development will continue to appear increasingly convincing and plausible in its responses, but in fact remain unwed from problem accuracy. The consequence of this development is not just a direct impact on both (mis)use and higher error rates (via further misapplication), but also feeds-back to earlier misplaced trust and oversight decline (C2 and C3), and feeds forward to uptake *breadth* compounds (C5).

It is worth reiterating assumptions regarding AI development timelines and the advent of AGI-like systems in relation to this compound. Specifically, from *current* through *transitional* states of AI development, towards AGI-like systems, we argue that RUUT risks *may* gradually decrease in terms of RUU inaccuracy (particularly relative to human performance), but that this may be offset by the overconfidence and over-application trends associated with cheaper, more persuasive developments. Furthermore, given above arguments regarding the *NP-Hardness* of even approximately optimal RUU - and particularly in the face of *deep uncertainty* inherent to *wicked* problems, combined with the ambiguity surrounding what form AGI-like systems will take, we argue it is by no means guaranteed that even the advent of an AGI-like "end state" will absolve the risk of errors. Critically, given there is no guarantee of development taking a "straight-line" in RUU accuracy improvements, as each time-point in that period still incurs the RUUT compounds, we should not assume this risk to simply resolve itself over time. Lastly, it is important to emphasise the distinction between potential development absent clear awareness of RUU performance issues (and methods of improvement described in Chapter 6) - which precipitate the RUUT, and one in which awareness and

> ***Example***: *As AI development continues, newer versions provide increasingly convincing policy recommendations as analysts in Organisation A continue to use such tools. This development reduces organisational costs and improves usability, directly increasing use, and (over)application, increasing error rates (e.g., "it now seems capable of handling XX…"). Analysts in Organisation A become more convinced (C2), and the need for human oversight seems weaker (C3), and other, hold-out analysts gradually also become convinced to start using (C5).*

interventions have successfully been applied and guide development.

### 4.2.5. Widespread Uptake (C5)

Use spreads as norms change and fears of getting left behind increase.

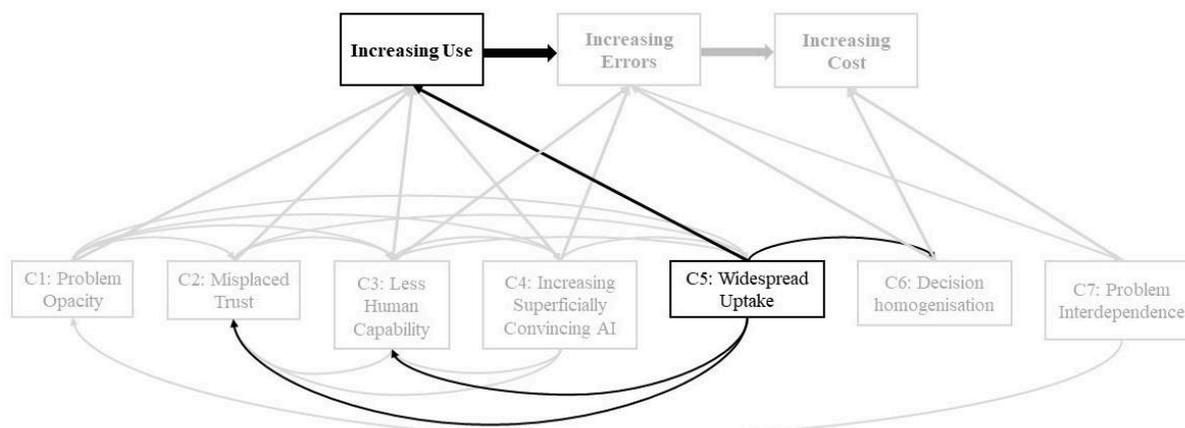

*Fig. C5. Map of Reasoning Under Uncertainty Trap (RUUT), highlighting Widespread Uptake compound (C5). Feed-forward effect to further compounds highlighted in arrows arching forward from top of C5 box, and feed-back effects arching backwards from bottom of C5 box, with direct impacts on central problem highlighted with straight arrows.*

Distinct from general use increase, which covers e.g., individuals applying AI tools more often to their problem spaces, the widespread uptake (C5) compounding factor covers increasing breadth of use across individual users, organisations, institutions, and problem spaces. Without clear feedback and error detection/correction (C1), as (misplaced) trust builds (C2), human RUU capabilities wither (C3) and AI tools appear to generate richer, more plausible outputs (C4), more and more prospective users across and within problem spaces will reach thresholds of 1) awareness and 2) sufficient incentives to also engage in AI RUU tool use. This is also driven by social/cultural and economic incentives, such as increasingly established new norms, and fears of losing competitive advantages. As described in Chapter 2, accuracy-related trust is not the only incentive that may induce widespread uptake. Although it could be argued that the absence of clear performance improvement indicators may slow uptake (i.e., trust cannot build), we note that the lack of erroneous performance indicators, coupled with cost, time, and social/cultural pressure incentives may be sufficient to induce widespread uptake.

> ***Example***: As analysts in Organisation A continue to use LLMs for their forecasting needs, unchecked by corrective feedback (C1), widespread use is both fed by misplaced trust and insufficient oversight, but also feeds back to signal more trust (C3) and foster further reliance (C3) - particularly as LLM responses become more plausible (C4). This incentivizes both hold-outs within Organisation A, and other organisations to adopt the seemingly improved tools (or be left behind). This spreading use feeds the RUUT directly, and leads to system-wide vulnerability (see C6).

Whilst this directly impacts (mis)use rates, such a compound has its own non-linear impacts as wider use further increases the aforementioned social/cultural and economic incentives, inducing potential tipping point behaviours. Further, wider use also acts as a signal feeding back into misplaced trust (C2) and lessening human RUU capabilities (C3), to further deleterious effect.

### 4.2.6. Decision Homogenization (C6)

Uniformity of (erroneously) advised decisions risks volatility.

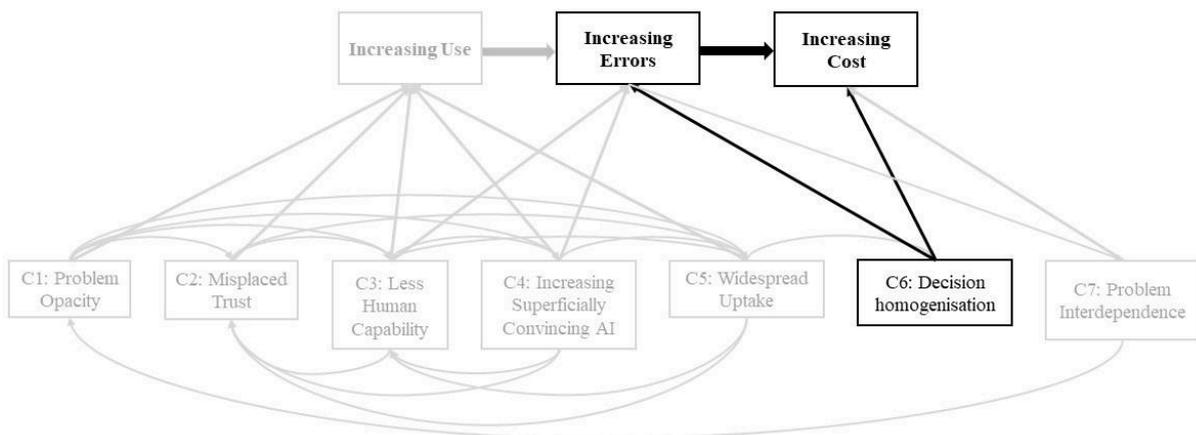

*Fig. C5. Map of Reasoning Under Uncertainty Trap (RUUT), highlighting Decision Homogenisation compound (C5). Direct impacts on central problem highlighted with straight arrows.*

As use becomes more widespread (C5), there is an additional compounding risk factor that comes into play: the homogeneity of decisions. Specifically, as more individuals, organisations, and institutions begin to rely on AI tools for the RUU basis of important decisions, the more the collective mass of decisions being made across an industry / nation / region share the same "advisor" basis, and thus arrive at the same/similar eventual decisions. Although such a top-down shared dependency *could* be beneficial *if* the shared advisor in questions is a) accurate, and b) is aware of the wider, coordinating picture (i.e., the advisor is aware of the potential collective impact from other advice-takers), this is far from guaranteed (see Chapter 3). As a consequence, this shared dependence (and resulting decision-homogeneity) becomes a substantial vulnerability. There are two primary reasons for this:

First, the more actors adopting the same, erroneous course of action, the less likely an alternative/exploratory, better course of action will be tested/demonstrated. We take pains to note here a key caveat - the homogenising dependency effect is reduced (but not eliminated) if users are effectively tailoring their generation of outputs (e.g., by varying prompts, tuning models). However, these tailoring processes are subject to effortful engagement, knowledge/training, and mixed incentives (e.g., seeking an "edge" vs time and efficiency losses) - which will suit certain industries/institutions/organisations more than others (e.g., tech-savvy industries, vs local public policy makers). We note that raising awareness and providing clear, accessible guidance and incentives to ensure effective tailoring across sectors is a worthy method of intervention, but we draw a distinction between the world in which this awareness is lacking (the current, RUUT issue), and one in which effective solutions are in place. Additionally, even with full awareness, we must still acknowledge that a common model architecture (and/or training data sources) will still foster *some degree of homogeneity*.

> ***Example***: *With multiple organisations now using the LLM tool for their forecasting, this "shared advisor" begins to inadvertently homogenise policy recommendations and decisions in those organisations, leaving open vulnerabilities to (now mutually) overlooked assumptions and mutual action. For example, organisation A in country A, and its opposite number in neighbouring country B are both recommended to acquire the same resource / area of land. All value it highly (and equally) and are therefore unwilling to back down, leading to misguided conflict.*

Second, the homogeneity of these (erroneously informed) decisions introduces potential volatility within a system. Stability within systems of actors often stems from the heterogeneity of those actors' preferences and actions (see e.g., Miller & Page, 2009). For example, the stock market remains broadly stable when there are both buyers and sellers of various stocks, willing to accept different prices from one another. Policy-makers debate a plurality of ideas that should represent the pluralities among the electorate, ideally reaching a representative compromise. Producers make different goods from one another based on differences in expectations of success / current and future markets. If these actors are all making the same decision about stocks, policies, products, etc., then substantial inefficiencies (at best) and crippling shocks (at worst) can result. For example, shortages in needed products/materials that were overlooked by AI RUU processes, policies that harmfully interact when engaged in aggregate across nations (e.g., waste exporting), and so on. Potential harms from this volatility, especially when they impact infrastructure and culture can have long-lasting harmful impacts, beyond the immediate harms (e.g., localised famine in the short term, damaged environment/soil in the long term).

This compound thus directly impacts not only the potential number of errors made with (mis)use, but also the potential costs of these errors. This latter aspect feeds forward to the final compound: *Problem Interdependence*.

*4.2.7. Problem Interdependence (C7)*

Siloed "solutions" cascades costs to interlinked problems.

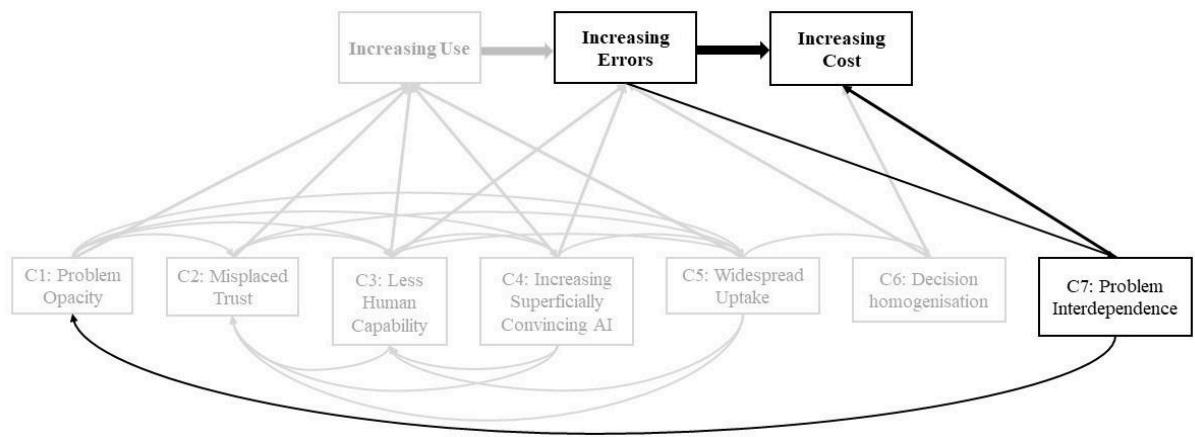

*Fig. C7. Map of Reasoning Under Uncertainty Trap (RUUT), highlighting Problem Interdependence compound (C7). Feed-back effects highlighted in arrow arching backwards from bottom of C7 box, with direct impacts on central problem highlighted with straight arrows.*

Finally, there is a compounding factor that directly impacts the cost/magnitude of errors. In essence, as the problems being "addressed" via AI RUU capabilities are not necessarily isolated - an aspect of their *super wickedness* (Levin et al., 2012), bad decisions in one area can have non-linear collateral damage in others. Damage in these problems can cascade outward, like a stone cast into a pond. For example, an AI tool may propose a policy that *should* solve a food security concern over the next 10 years (e.g., investing in particular technologies and seeds for a crop-type that should withstand expected climate and soil conditions). However, this policy does not account for e.g., aggregated pollutants in downstream bodies of water, reliance on a form of oil/plastics, or food producer livelihood effects (and a subsequent farmer exodus), each of which may then contribute to further environmental or social/societal unrest/harms. This problem is particularly likely when AI tools are used for RUU tasks where *assumed scope* (or alignment) is more limited than the user realises.

> ***Example***: *As Organisation A continues to use the tool for forecasting within their scope/remit - prompting the tool within that scope, recommendations appear well-tailored, but fail to account for interactions with other remits (both nationally and internationally). For example, a recommendation to isolate from misplaced belief in an upcoming pandemic instead results in an eventual depletion of a food stock component that relied on trade. These semi-blind interventions may only make the difficulty of predicting future states worse (C1), and raise the societal-scale risks.*

This compound not only directly impacts the cost of potential errors, but also feeds back to the original compound: problem opacity (C1). Given (erroneous) interventions/actions in one area may have unintended (in)direct impacts in another area, this makes getting an accurate picture of current and future states of any given area increasingly difficult as a function of each additional action taken.

### *4.3. Summary: RUUT and Timelines*

Considering the compounds that comprise the RUUT as a network of interlinked structural risks, we can consider their combined effects by returning to an approximate timeline of (relative) expected decision utility. Crucially, as the linkage between compounds primarily consist of positive feedback loops, the accelerative impact is likely to be nonlinear (Forrester, 1961).

    Consequently, absent corrective intervention, we can expect a non-linear decrease in the expected utility derived from the continued use of AI tools for RUU-based solutions to wicked problems. Specifically, the compounds described not only lead to continually increasing use at a non-linear rate, but also non-linear increases in the likelihood of those uses being erroneous, and the potential costs of those errors. The rates of each of these increases, we argue, will accelerate over time as the feed-backs gain momentum.

    The compounds laid out in this chapter are by no means exhaustive. As we will touch upon in the next chapter, given the general purpose technology aspect of AI tools, there are likely other related ways in which the effects described here may be affected. Some may be foreseen (e.g., automation of science), whilst others will not be, regardless, we can assume the GPT nature of AI tools will entail more substantive (i.e., wide-spread / pervasive) impact across economies and societies (Bresnahan & Trajtenberg, 1995; Gruetzemacher & Whittlestone, 2022), irrespective of the degree of RUU capabilities inherent to those impacts.

Although such projections are highly uncertain given the number and complexity of variables involved - and thus it remains possible for positive trajectories - much of this uncertainty is derived from our lack of knowledge regarding AI RUU capabilities and their trajectory. Without the solutions described in Chapter 6 to assist in guiding development and use towards such a positive outcome, we remain needlessly at risk of a societal-scale risk that stems from a seemingly "typical" misuse error risk, but when considered within the context of surrounding structural risks (and compounds) can have outsized, non-linear impact (and thus a substantially greater imperative to detect and correct).

**Chapter 5: Interrelated Risks**

*5.1. Current, Structural, and Societal-scale risks*

The central purpose of this work is to highlight the RUUT as a current structural risk that stands to grow in its potential societal-scale impact as timelines extend. However, it would be remiss to suggest this risk, and the compounds that underpin it, is isolated from the wider spectrum of identified AI safety and risk concerns. In fact, to assume so would be to overlook the very purpose of taking a structural risk perspective: that many risks are not only a function of myriad interrelated factors (e.g., chains of events), but also may be interdependent themselves. Here we describe several close AI risk topics as illustrative examples that relate closely to either AI RUU functionality directly, or affect (and are affected by) the underpinning compounds described in Chapter 4. We take pains to note that these related risks are by no means exhaustive, nor are they bound to specific timeline or impact scale assumptions.

*5.2. RUU ignorance and Long-Horizon Planning*

Blindness to true RUU capabilities in current tools lies not only at the heart of RUUT risks, but also dovetails with the transition towards an acknowledged longer-term risk. Long-horizon planning (LHP) has been argued as a potential AI capability with societal-scale potential consequences (Shevlane et al., 2023). Put simply, LHP requires RUU functionality. Defined as a capability to make and adapt long-term, multi-step plans across multiple domains and *without recourse to trial and error*, LHP is an enabling condition for AI existential risk stemming from misalignment.

    LHP therefore entails planning into the future *effectively* - requiring a capacity to predict that future, and the prospective impact of possible actions. Notably, this includes novel domains and scenarios, and therefore places it firmly as a reasoning under uncertainty task (i.e., making accurate inferences about future events from incomplete/uncertain knowledge in the present).

    Linked to the compound described in Chapter 4 regarding deleterious AI development trajectories, without due scrutiny (i.e., thorough evaluation, benchmarking, and oversight initiatives - discussed below) into RUU capabilities, potential LHP capabilities that are based on RUU remain effectively undetectable. Whether an AI tool is *bad at RUU* (lowering LHP risk, but increasing RUUT), or is particularly *good* (reducing RUUT, but increasing LHP risks), remaining oblivious raises risks unnecessarily. Absent intervention (see Chapter 6), and particularly absent awareness, we may unknowingly occupy either of two possible worlds: One in which the AI RUU performance does not improve beyond human performance, and thus RUUT results, with accumulated negative impacts building in a non-linear fashion. The other, in which AI tools do manage to circumvent this **hardness** and gain LHP capabilities (or, given potential hard limits in this capacity, see Heninger & Johnson, 2023; at least improve beyond human performance and detection), such that potential societal gains from such a capable tool may be entirely subsumed by the misalignment societal-scale risk associated - a risk that outweighs potential gains the fast current AGI timelines trend.

    Put simply, the parent cause of RUUT dangers - the ignorance of RUU (in)capabilities (both among users, but also among developers) - is also a parent cause of unchecked societal-scale risk in the long-term. Critically, this is *regardless* of where one stands regarding the capacity of where prospective AGI or AIXI (Hutton, 2007) / superintelligences (Bostrom, 2014) lie in resolving the **NP-Hardness** (and thus accuracy gains) of RUU functionality.

*5.3. The Structural Risk of the Automation of Science*

Within AI research, the automation of science is based on the notion of "self-guided" tools or instruments that can perform scientific research in the absence of human intervention (see e.g., King et al., 2009). Whilst historically this has been considered within the context of specialised tasks and topic areas (e.g., genome sequencing, Dias & Torkamani, 2019; protein-folding, Callaway, 2022; but see Moore et al., 2022), wherein computations of permutations are a core automated skill, more recent discussion has turned to more generalist scientific capabilities. These latter capabilities include devising hypotheses, testing these hypotheses against available (or generated) data (i.e., devising suitable *tests*), analysing and reporting the findings of said tests, and subsequently revising hypotheses accordingly (King et al., 2009).

It is the successful AI implementation of the hypothetico-deductive method, which entails finding the most parsimonious explanation of current facts in such a formulation as to be falsifiable with subsequent data gathering, that is of most interest to the current work. Such a process entails reasoning under uncertainty: devising a hypothesis necessarily is based on incomplete/imperfect information, from which a prediction may be generated. Whilst recent work on inductive inference capabilities of LLMs reveals a strong capacity for hypothesis generation and refinement, this requires strict task rules (Qui et al, 2023). It is where the strictness of the task/problem at hand is less apparent that we argue RUUT-related vulnerabilities may appear. Consequently, automation of science may be affected by the issues described here in the following ways:

- **Problem Opacity**. Although incremental scientific enquiry should entail principles of occam's razor and falsification in hypothesis building, testing, and revising, there remains a risk of problem opacity and the accompanying difficulties of RUU (see Chapter 3) impacting the validity/accuracy of the (automated) process.
- **Misplaced Trust and Weakened Oversight**. In a similar fashion, as automation of science begins to show signs of apparent progress - some of which may in fact be sound, valuable discoveries - trust in this automation and its outputs grows. With this growing trust (and the lack of awareness as to when it is misplaced), comes the accompanying compound of weakening human oversight: although empirical tests may mitigate this issue, in more complex and theoretical domains, the need for correction is less apparent, and thus the incentive to place effort into human oversight diminishes, reducing the capacity to uncover "cracks in the knowledge foundation" even further.
- **AI tool development (convincingness vs accuracy)**. As with RUUT, the path of AI tool model development is not guaranteed to be married up to accuracy improvements. Whilst testable improvements in automated scientific processes are a marketable (and therefore incentivized) goal for developers, the degree to which these improvements extend towards harder, more RUU intensive and wicked problems is a) less detectable via current evaluation processes, b) not guaranteed from more straightforward reasoning gains (see Chapter 3), and therefore c) whilst potentially seeing gains in apparent *convincingness*, may in fact engender further misplaced trust in automation of science in such cases.

From the structural perspective, we should also note the ways in which automation of science may in turn affect RUUT:

- **Evidence Base Generation: Vulnerability**. Given the scientific intention to build upon current knowledge, if that knowledge is in fact flawed in an unknown way, then compounding errors may result from use of the expanding knowledge base, spreading to other problem domains (notably those of wicked problems that intersect with scientific advice, e.g., sustainability problems). An already limited evidence base is particularly vulnerable to the influence of new entrants or (unrealised) poor quality; enhancing the deleterious impacts of RUUT via the central error rate increase from (mis)use effect.

- **Evidence Base Generation: Homogeneity**. Risks regarding increased homogeneity in knowledge production and creativity have been well documented in terms of "algorithmic monocultures" (Doshi & Hauser, 2023; Kleinberg & Raghavan, 2021; Toups et al., 2023). In the particular context of RUUT, this feeds into the homogeneity problems highlighted in Section 4.2.6 (Decision Homogenisation) - further increasing the risks of collective poor decisions or overlooked solution areas, critically, as the collectively poor decisions cannot be assumed as independent in their effects on the broader system, whether via volatility (Section 4.3.6) or via catalytic/cascade/nonlinear interactions (Section 4.2.7).

As we will describe in Chapter 6, fortunately many of the solutions for RUUT can directly assist automation of science risks. Moreover, unlike some of the wicked and super wicked problems that can drive RUUT, automation of scientific progress is not restricted by the same deleterious pressures and opacities. Specifically, the purpose of scientific enquiry is to devise increasingly more accurate models of the world. The nature of this process is to test hypotheses against data, and thus through error/falsification (i.e., *feedback*) reject inaccurate beliefs. Consequently, with strict application protocols for the automation of science, hypothesis generation and refinement is not necessarily error prone (see e.g., Qiu et al., 2023). Our purpose here has been to highlight the susceptibility (and potential deleterious impact) of the automation of science to RUUT-like compounds and effects (and vice versa), if deployed without due care to opacity and wickedness concerns.

## Chapter 6: The Solution Space

*6.1. No silver bullets*

The Reasoning-Under-Uncertainty Trap (RUUT) presents a challenge of considerable complexity, situated at the confluence of human actors, the systemic environments they operate within, and the capabilities and limitations of AI tools. Addressing the central issue of AI tool misuse in RUUT demands an appreciation of the intricate web of structural risks and interdependencies that characterise this landscape. The misuse of AI tools in reasoning under uncertainty (RUU) scenarios is not merely a problem of user error or a reflection of deficits in AI development. It's a symptom of a broader, more systemic issue. This systemic issue stems from a complex interaction between users, their beliefs, the reinforcement of these beliefs by organisational incentives, and the unique properties of the problem domains they engage with. These elements do not operate in isolation but are part of an intricate feedback loop that shapes, and is shaped by, the use and misuse of AI tools.

When considering solutions, it is critical to not only address the central RUU misuse risk, but also look beyond it. The structural risks, encompassing feedback loops and interactions at various levels, necessitate interventions that address not just the surface symptoms but the underlying causes. This perspective aligns with recent sociotechnical approaches to AI safety, which advocate for a holistic view encompassing both social and technical dimensions (Lazar & Nelson, 2023; Weidinger et al, 2023). Complex systems theory offers valuable insights for identifying sensitive intervention points within the RUUT context (Farmer et al., 2019). In complex systems, change can be affected not just by direct interventions but also by altering system-wide properties or goals (Meadows, 2008). This approach suggests that solutions to the RUUT problem should be multi-pronged, addressing different aspects of the issue in a coordinated manner.

From a structural risk perspective, it becomes clear that solutions must encompass a broad spectrum of interventions. These may range from addressing individual cognitive biases and improving AI tool design to altering organisational incentives and reshaping broader socio-technical systems. The recent push in AI safety research advocates for such comprehensive approaches, recognizing that the effective management of RUUT requires interventions at multiple levels – individual, organisational, and systemic. Given the complexity and multi-faceted nature of the RUUT problem, this chapter proposes a three-pronged approach to solutions:

**Awareness**: Raising awareness of the RUU weaknesses inherent in AI tools, focusing on both users and developers, which simultaneously mitigate specific compounding risks.

**Investigation**: Rigorous investigation into the RUU capabilities of AI tools, including the development of robust evaluations, metrics, and benchmarks, and empirical comparisons with human performance across various domains.

**Intervention**: Exploring and implementing methods to improve RUU in AI tools, such as prompt engineering and training set enhancements, while acknowledging and addressing the fundamental difficulties posed by combinatorial explosiveness and deep uncertainty.

In summary, addressing the RUUT problem requires a nuanced and multi-faceted approach, one that recognizes and targets the complex interplay of factors contributing to this challenge. By combining awareness, investigation, and intervention strategies and grounding them in a structural risk perspective, this approach aims to mitigate the misuse of AI tools in RUU and foster safer, more effective decision-making processes within organisations. This comprehensive strategy aligns with the latest thinking in complex systems theory and sociotechnical approaches to AI safety, promising a more holistic and effective pathway to addressing the intricacies of RUUT.

*6.2. Awareness*

Awareness of the inherent weaknesses of AI tools in reasoning under uncertainty (RUU) stands as a foundational step in mitigating risks and guiding the development and use of these technologies. This awareness spans across various domains: from AI developers and researchers to end-users in individual and organisational settings, and finally to governance bodies responsible for policy-making (e.g., government oversight, such as US House and Senate Intelligence Councils; as well as regulatory bodies and agencies, such as the NSF's Directorate for Computer & Information Sciences and Engineering (CISE)).

For developers and researchers, awareness of AI's limitations in RUU is not just an abstract concern; it is a prerequisite for meaningful progress. Recognizing these weaknesses is essential for developing effective evaluation methods and benchmarks that genuinely reflect an AI system's capability to handle uncertainty (as described in Sections 6.3 and 6.4 below). It can also directly mitigate the problematic AI development compound (C4), such that risks of misuse are not only addressed directly, but also via feedback loops to growing misplaced trust (C2), weakening oversight (C3), and widespread inappropriate uptake (C5).

At the user level, heightened awareness serves multiple purposes. It helps temper expectations, and directly begins to address the central inappropriate demand for AI tools in the RUU domain. Further, it helps ensure that users do not place undue trust in AI systems (C2), a factor that could otherwise lead to weakened oversight (C3) and potentially dangerous decisions. Furthermore, by understanding the limitations of AI in RUU, organisations can make more informed decisions about adopting and integrating these tools, potentially slowing the rate of widespread, uncritical uptake (C5).

For governance bodies, awareness is a cornerstone for informed policy-making. It enables the creation of regulations that ensure transparency in the use of AI tools within public bodies. However, this transparency needs to be coupled with best practices and a commitment to accuracy; otherwise, it risks accelerating misuse rates. Historical precedents show that transparency initiatives can effectively mitigate and even reverse harmful incentive traps and feedback loops (Farmer et al., 2019). A system-wide emphasis on awareness can gradually shift the overarching goal towards prioritising accuracy. This shift is a powerful lever, as it influences top-down actions across the system, including actor behaviour and structural incentives (Meadows, 2008). When accuracy becomes a central goal, it cascades through the entire system, reshaping incentives, decision-making processes, and ultimately, outcomes.

The role of awareness in tackling the RUUT problem cannot be overstated. It acts as a multiplier of positive change, setting the stage for more nuanced interventions and better-informed decisions at all levels. By fostering a deep understanding of the limitations and potential risks associated with AI in RUU contexts, awareness becomes the first crucial step in a series of interventions aimed at mitigating the challenges posed by the Reasoning-Under-Uncertainty Trap.

### 6.3. Investigation

A comprehensive approach to addressing the central risk within the Reasoning-Under-Uncertainty Trap (RUUT) necessitates a thorough understanding of the current capabilities of AI tools in handling RUU. While significant strides have been made in evaluating and benchmarking AI capabilities in causal (Jin et al., 2023; Tang et al., 2023), arithmetic (Imani & Shrivastava, 2023), commonsense (Marcus & Davis, 2020), and logical reasoning (Creswell et al., 2022) capabilities, these assessments do not adequately capture the nuances of RUU processes.

RUU is not easily measurable, given the inherent uncertainties and combinatorial explosiveness. However, there are existing evaluative reasoning standards that could be applied, such as coherence-based frameworks (e.g., Glass, 2002; Schum, 1994)[14]. Critically, this form of evaluation will require substantial work in the devising of gold standard datasets of problems, which target a range of core RUU sub-structures and uncertainty types, and in the creation of corresponding rubrics that can be used in qualitative (human) assessments of outputs. This will be laboursome, given the breadth of possible combinations and problem contexts, and so understanding where automation/hybridisation can be introduced is a valuable area of enquiry. Further, as will be noted in the interventions listed below (Section 6.4), processes which improve transparency of model reasoning (e.g., mechanistic interpretability improvements via expansion of chain of thought methods, such as tree of thought, see e.g., Yao et al., 2023) should also be a priority if we are to understand how AI tools are navigating underlying complexity.

A critical area of investigation is understanding how increasing combinatorial explosiveness and deep uncertainty impact AI's performance on RUU tasks. Consistency in handling these complex problem types will be a key metric of evaluation. Efforts like the Eliciting Latent Knowledge protocols (Christiano et al., 2022) represent steps in this direction but would require expansion to cover the specific challenges posed by RUU. Solutions to these challenges will benefit from incorporating measures of decision-quality taken from the substantive foundations already laid down by DMDU research, such as shifting from trying to determine the most accurate decision as decision-spaces expand, to a focus on the simulated robustness scores of possible decisions in relation to desired outcomes *across the space* (see RDM, Lempert et al., 1996; 2003; 2013).

Although a potential solution to determine RUU accuracy could be to back-cast models against historical RUU / forecasting problems, given the deep uncertainties and uniqueness of the wicked problems being addressed, there is the obvious problem of sufficient data / example availability (i.e., a practical impossibility failure, Raji et al., 2022). This raises questions of the external validity of outputs, and the internal validity of (in)consistent underlying reasoning, given the varying degrees of uncertainty inherent in such problems. The data-impoverished nature of *wicked* problems poses a significant challenge to training AI systems with generalist capabilities in RUU. This reality underscores the need for self-policing and the effortful nature of this type of evaluation within the broader context of structural risk.

Given the crux of the RUU problem is the relative performance of AI tools to human counterparts, it is also imperative to continue empirical investigations into comparing AI tool RUU performance with human performance across multiple problem types and domains. These, much like the design of gold standard problems and rubric sets, should encompass a broad spectrum of problem complexities and deep uncertainties, providing a more holistic view of AI's competencies and limitations in RUU.

---

[14] Our argument here is a focus on improving the application of evaluative standards, which not only entails the application of existing appropriate standards, but also the investment in new standard creation. We see this as akin to the call for evaluative standards by the US Office of the Director of National Intelligence (US ODNI, 2015) in recognition of past overlooked (and deleterious) reasoning shortcomings (Robb et al., 2005). In this regard we note the RUUT may amplify the breadth and depth of overlooked shortcomings without such a return to evaluative scrutiny.

Finally, the compounding risks described in the RUUT problem also highlight a gap in current evaluations regarding structural risks and their potential impacts, which have recently started gathering attention under the banner of socio-technical safety evaluation (Lazar & Nelson, 2023; Weidinger et al., 2023). Addressing this gap more broadly could help address potential underestimations of potential AI risks more broadly, as well as expose more productive points of intervention. This is a novel, hard problem. Evaluating potential complex interdependencies in human-human and human-environment systems is notoriously challenging and laboursome - and now even more so given expansion to cover human-machine and machine-environment interactions. However, when we are attempting to evaluate structural risks, we can look to pre-existing work from the field of systems thinking.

Systems mapping is an evaluative process for determining how parts of a system are connected, the causal relationships between those parts, and the insights that can be derived from their interactions (see e.g., Barbrook-Johnson & Penn, 2022). Relevant approaches include Bayesian Belief Networks, Causal loop diagrams, and system dynamics models (stock and flow). These approaches, which will rely on substantive subject matter and systems dynamics / complex systems expertise are essential to gaining a better understanding of the potential interconnections within a system (and subsequent emergent and non-linear behaviours). Further, they provide a roadmap of potential points of leverage, and thus enable the development of strategies for more efficient and effective mitigation (or enhancement in the case of positive outcomes). As a method of evaluating AI models, this is going to be intrinsically human-led, as mapping is primarily based on elicited human judgement of prospective connections. Nevertheless, hybrid methods should be explored (e.g., integration with causal discovery methods) given the magnitude of the potential evaluative space. Critically, such an evaluative approach by necessity entails the integration of multiple perspectives (see arguments made in e.g., Lazar & Nelson, 2023), necessary to maximising the likelihood of catching notoriously challenging effects (i.e., structural risks, and their interconnections). This is particularly essential given the temporal dynamics of such effects - snowballing effects must be diverted or stopped before they rapidly acquire size and momentum.

In conclusion, investigating AI's capabilities in RUU demands a multi-dimensional approach that extends beyond traditional evaluation metrics. This approach must encompass the creation of comprehensive datasets[15], the development of adaptable evaluative rubrics, and the consideration of structural risks. Only through such a holistic evaluation can we hope to address the central risk within the RUUT and leverage AI's full potential in complex decision-making environments.

### *6.4. Intervention*

There are two main branches of intervention to consider with respect to RUUT: improvements in AI development to enhance RUU functionality and the formulation of robust AI safety policies. Both branches are crucial for mitigating the risks associated with AI tool misuse in complex decision-making scenarios.

---

[15] Namely, datasets that expand beyond state action spaces, to incorporate interactions with surrounding problem space characteristics (e.g., via enriched annotation of identified problem reasoning structures, missing/partial information sources and types) and user directions (e.g., user prompt features, such as target decision properties like accuracy vs robustness).

With the development of suitable evaluations and benchmarks (Section 6.3), determining potential inroads for RUU improvement in AI tools becomes possible. However, we must not lose sight of the potential limitations on improvement - the NP-hardness of combinatorial explosiveness and deep uncertainty inherent to RUU - when determining appropriate ambition and expectations of performance improvements. Instead, and sufficient to head off key RUUT concerns, there is an imperative to develop and confirm reliable improvements over human performance.

In the context of AI development avenues, there are a number of potentially fruitful avenues for exploring potential RUU improvements (given aforementioned work on sufficient measurement), many of which we note are not only independently additive, but mutually assistive:

**Prompt Engineering**: For RUU tasks, prompt engineering can involve designing queries that explicitly probe for underlying reasoning processes, and assess appropriate and consistent accounting for uncertainties and ambiguities. For example, expansion of chain of thought reasoning probes to encompass more generalised "tree of thought" prompting have begun to yield gains in novel planning tasks (Yao et al., 2023). Another fruitful avenue of guidance for prompt engineering can be found in the DMDU literature. Specifically, key insights from the study of how to make better, higher quality decisions when faced with deep uncertainty involve knowing the correct questions to ask (and correspondingly, the more appropriate metrics against which to judge outputs (for a review see Marchau et al., 2019). Setting appropriate task constraints and metrics is key to salvaging performance, and in this case avoiding higher likelihoods of hallucinated (erroneous) responses that stem from having asked the *wrong question* (e.g., shifting away from predictive accuracy to robustness, Lempert et al., 2003).

**Hyperparameter Tuning**: Adjusting the hyperparameters of AI models is a necessary method for seeking out possible optimisations of problem-domain to response output accuracy and consistency, particularly given available data (and conversely, remaining uncertainty) are central to hyperparameter tuning considerations, and inherent to RUU problems/tasks.

**Developing Specialized Training Sets**: Given the aforementioned need for the creation of gold standard datasets for evaluation and benchmarking purposes (Section 6.3), the skills and labour required also lend themselves to training set developments that can assist in training AI models to better handle RUU scenarios. These datasets should include a wide variety of problem types and levels of uncertainty.

**APIs and User Training**: The development of APIs and user training initiatives can not only independently reduce potential misuse risks (e.g., improving understanding of appropriate RUU performance expectations), but can also be leveraged to address more troublesome aspects of RUU tasks, including more explicit consideration of combinatorial explosiveness and deep uncertainty influences (e.g., APIs that can assist in estimating problem "size", and/or provide subsequent iterated probes of posed problems and questions to help illustrate the impact of residual uncertainty).

More concretely, when it comes to improving the quality of outputs, given the end-goal is better quality decisions, we can again turn to the considerable work in DMDU to develop approaches that address the *hard* challenges associated with RUU problems described here. In this particular case, APIs could be developed that integrate one or more DMDU approaches to best leverage these insights for more effective model outputs - as they pertain to not only the user's needs, but also the hard realities of the problem spaces. Some examples include (but are not limited to):

*Robust Decision Making (RDM)*: Shifting the focus from trying to make better predictions, to yielding policies/decisions that perform well across a range of potential scenarios, via a combination of techniques, including Assumption-Based Planning, scenarios, and Exploratory Modelling (Lempert et al., 2003). Robustness of model outputs is already of interest to AI research and development, this can be considered an extension of the same principles.

*Info-Gap Decision Theory (IG)*: Identifying disparities between known and potential unknowns as they pertain to the robustness of policies to failure. Such a technique is computational in nature, evaluating actions in a non-probabilistic fashion (Ben-Haim, 2006). Actions and alternatives (the basis of this analysis) could be generated through an API / series of prompts, incorporating both human and AI tool generation. Identified gaps could then be fed into further decision-approach processes, or at minimum made clear to users.

*Dynamic Adaptive Planning (DAP)*: Focusses on adaptation of plans and contingencies to changing/updating conditions - split into design and implementation phases, both of which entail a monitoring component (Walker et al., 2001). Within the context of AI RUU APIs, such an approach can be integrated into response structuring, tagging, and revision components, with monitoring thresholds being maintained by separate agents (human. Machine, or both conceivably).

In general, the integration of these approaches into an RUU focussed API will boost decision quality (and therefore remediate the RUUT) via 1) identification of problem-relevant combinatorial explosiveness / deep uncertainty factors (e.g., expansive decision-option sets; shifting/changing timelines), 2) application of more appropriate metrics for deriving relative quality among decisions (e.g., robustness), and 3) appropriate outputs that minimise risks of overconfidence (i.e., via inadvertent exposure / corruption from unaccounted deep uncertainty influences). For a relevant review of DMDU approaches, see Marchau et al (2019).

**Leveraging Advanced Hardware**: Outside of work with existing models and hardware, one obvious consideration is the direction and speed of hardware development. As mentioned previously (see Sections 3.3 and 3.4), development of quantum computing can offer greater returns on handling degrees of necessary compute dictated by the combinatorial explosiveness of the RUU problem in question. Of course, code-bases, softwares, and models that can better leverage these hardwares (current and prospective) could further garner meaningful gains in performance via broader coverage of possible worlds.

In line with our structural / socio-technical perspective for addressing the risk of RUUT, there are additional "levers" that should be considered beyond the development of AI tools. Specifically, formulating and implementing effective AI safety policies is critical, as it can impact not only classes of actors in the system (developers, users) directly, but also shift system goals and incentives for more efficient alignments. We outline several areas of AI policy as potentially fruitful avenues:

**Accountability in Usage**: Just as scientists are expected to detail their methods and tools, there is value in users of AI tools for critical decision-making scenarios being held accountable for how they use these tools. This includes documenting the AI models used, the rationale behind their choices, and acknowledging the limitations of these tools.

**Disclaimers and Flags**: To not only enable accountability initiatives, but also ensure incentivisation appropriately reaches developers, we suggest there is value in mandating the attachment of bolstered disclaimers or warning flags to outputs generated by AI tools, especially for functions prone to inaccuracies. This practice (though present in limited, voluntary degrees) will help users remain cognizant of the potential errors and approach the information with the necessary scepticism, but also continues to incentivise accuracy-based development (the caveat here remaining effective external corroboration and evaluation of tool performance).

**Broad Perspectives in Oversight**: In line with recent calls (Lazar & Nelson, 2023; Gruetzemacher et al., 2023), AI safety and oversight should involve diverse expertise, backgrounds, and perspectives. This diversity is essential to identify and mitigate risks that may not be apparent from a single perspective, and reduce the potential influence of compromised incentives (e.g., internal self-policing).

**Adaptive Foresight in Policy Frameworks**: We have also raised the issue of *wicked* problems, of which AI safety and governance is one. The immediate policy imperative is to adjust the policy framework towards adaptation/dynamism, such as 'adaptive foresight' approaches (Eriksson & Weber, 2008).

In addressing the RUUT problem, interventions must span both the development of AI tools and the policies governing their use. On the development front, continuous improvement in AI's RUU capabilities through technical advancements and training is paramount. Concurrently, AI safety policies need to emphasise accountability, transparency, adaptability and broad structural perspectives, to ensure not only that these tools are used responsibly and effectively, but that potential associated structural risks are not only detected, but accounted for. This dual approach, combining technical innovation with robust policy frameworks, offers the most promising path to enhancing AI's capacity for reasoning under uncertainty and ensuring its safe and beneficial application.

*6.5. Summary: Correcting the Timeline*

This chapter has outlined a prospective strategy to address the Reasoning-Under-Uncertainty Trap (RUUT), highlighting the need for a multi-pronged approach to tackle the complex and structural risks inherent in the misuse of AI tools in RUU problems. The solution to the challenges posed by RUUT lies not only in the technical development of AI but also in considering the broader structural aspects involving user behaviour, governance, and the AI development community. Specifically, we have outlined 3 categories of solution, each building off the last, but of substantive independent worth:

**Awareness**: The first critical step in mitigating the risks of RUUT is raising awareness among developers, users, and governance bodies about the limitations of AI in reasoning under uncertainty. This awareness is foundational for setting realistic expectations, guiding responsible usage, and informing policy-making.

**Investigation**: Following awareness, the investigation phase involves developing robust evaluation and benchmarking tools to accurately assess the current RUU capabilities of AI systems. This step is crucial for identifying specific areas for improvement and for understanding how AI tools handle complexities involving uncertainty and decision-making, and the risks relating to relative performance against human counterparts.

**Intervention**: The final step in addressing RUUT is intervention, which includes both technical enhancements in AI development and the formulation of effective safety policies. This involves improving AI's RUU functionalities through methods like prompt engineering and hyperparameter tuning, as well as implementing policies that ensure accountability in AI usage, and smart adaptability in AI safety policy.

Collectively, these initiatives—awareness, investigation, and intervention—have the potential to counteract and possibly reverse the harmful compounds within RUUT. By addressing both the technical and structural aspects of RUUT, these steps can lead to safer, more reliable, and effective use of AI tools, and serve as an example for dealing with structural risks.

An overarching theme in this approach is the necessity for identifying and exploiting sensitive intervention points in policy-making (Farmer et al., 2019). By identifying the compounding effects and feedback loops within structural risks, we enable more effective targeting of more impactful influences over desired goal states. Given the issue of AI safety is ultimately a *wicked* problem, tackling this requires not only a deeper, broader understanding of risk factors, and how they interrelate, but also an adaptive and dynamic approach to policy formulation and implementation. The concept of 'adaptive foresight' serves as a valuable example in this regard (Eriksson & Weber, 2008). It suggests that policy frameworks should be flexible and capable of evolving in response to new information and changing circumstances. This adaptability is essential in the fast-evolving domain of AI, where new developments can rapidly alter the landscape of risks and opportunities.

In summary, the journey towards mitigating the risks associated with RUUT and safely navigating the development and application of AI is intricate and multi-dimensional. It necessitates a concerted effort across various domains, involving technical advancements, enhanced awareness, rigorous evaluation, and dynamic policy-making. By identifying and exploiting sensitive intervention points across these areas, there is potential to steer the use of AI towards a trajectory that maximises benefits while minimising risks, ultimately contributing to a safer and more responsible AI-enabled world.

## Chapter 7: Conclusions

### *7.1. The RUUT problem*

The challenges facing humanity are immense and interconnected, demanding solutions to some of the most complex reasoning tasks ever encountered. Understanding and intervening in a future rife with uncertainty and complexity ranks among these formidable challenges. In an age where the capabilities of Large Language Models (LLMs) and other AI tools are rapidly advancing, there is an increasing temptation, especially among decision-makers who may not fully understand the intricacies of these technologies, to rely on them for addressing complex and uncertain problems. This reliance is not without risk. The allure of AI as a panacea for complex decision-making is a siren call, leading potentially to a landscape where critical decisions are made without fully grasping the limitations and implications of the tools at hand.

Here we have described a phenomenon that we term the Reasoning-Under-Uncertainty Trap (RUUT), which is a current AI risk that stems from the above demand/challenge. The problem manifests in two distinct but interconnected aspects:

**Awareness and Misuse Risks**: There exists a significant lack of awareness about the capabilities and weaknesses of AI tools in reasoning under uncertainty. This gap in understanding poses a substantial risk of misuse, particularly among users who may naively apply these tools to solve complex problems. As a result, we risk costly misuse errors that, given the opacity of the *wicked* problems at hand, we are unlikely to catch and correct. The crucial question is not only about the inherent limits of AI tools in RUU scenarios but also about ensuring that their performance is robust and reliable, especially when compared to human decision-makers.

**Systemic and Structural Risks**: The risks associated with the misuse of AI tools do not exist in a vacuum. They are embedded within a broader system or structure that is dynamic and responsive. Within this system, interlinked feedback loops can compound, exacerbate and accelerate the harms associated with this misuse. For example, as unchecked use leads to growing misplaced trust, incentives to retain (costly) human oversights diminish, and this weakened oversight sets new harmful norms where the already compromised capacity to detect and prevent errors is reduced even further.

The RUUT problem we describe here highlights how AI risks are not standalone issues, but are deeply embedded within a network of systemic interactions. We have illustrated how one central "component failure" (AI tool misuse for RUU problems) can interact with the dynamics of the complex systems users, problems, tools, developers, markets, and governance, inhabit, producing cascading effects across the entire system. The implications of RUUT as emblematic of a broader class of AI risk are profound and far-reaching. As decision-makers increasingly turn to AI tools to navigate complex scenarios, the potential for misuse and the resulting systemic risks rise correspondingly. This makes it imperative to address these forms of risk not just at the level of individual AI tools or users, but with an understanding of the broader systemic and structural dimensions at play.

In sum, the RUUT problem presents a scenario where the unexamined use of AI tools in complex decision-making could lead to significant and unintended consequences. It calls for a concerted effort to understand and mitigate these risks, emphasising the need for awareness, investigation, and intervention at multiple levels. As we navigate towards increasingly uncertain futures, addressing the RUUT problem becomes not just a matter of technological advancement, but a fundamental requirement for responsible and effective decision-making in a world shaped by artificial intelligence.

### *7.2. Solutions & Implications*

Addressing the RUUT challenge requires a multifaceted approach that begins with awareness and extends into targeted interventions in AI tool development and safety policy.

Awareness is the cornerstone of addressing the RUUT problem. Without a clear understanding of the risks associated with AI tool misuse, particularly in reasoning under uncertainty, decision-makers, developers, and users remain vulnerable to costly errors. Elevating awareness does more than just reduce the immediate risk of misuse; it actively counters several harmful compounds within the system, such as misplaced trust in AI capabilities and weakening oversight mechanisms. This approach aligns with the principles of Sensitive Intervention Points (SIPs; Farmer et al., 2019), and the leverage points in complex systems (Meadows, 2008). By enhancing awareness, we can initiate a chain reaction that mitigates broader risks and facilitates more informed decision-making. This is particularly essential given the temporal sensitivity inherent to non-linear systems (i.e., it is easiest to prevent/mitigate non-linear effects before they snowball), but even more so in a problem-space where the acting impetus (i.e., AI tool) is also subject to rapidly evolving timeframes.

Following initial, system-wide awareness, there are two interrelated paths towards not only mitigating RUUT, but addressing and improving RUU-related outcomes:

In the realm of AI tool development, there is substantive necessary work to expand current evaluation shortcomings. In particular, evaluating RUU requires a rigorous approach to creating gold standard datasets and rubrics, which cover a wide range of underlying problem structures, questions, and contexts. This approach is essential not only for evaluating AI capabilities but also for informing the design of effective training sets that can improve AI performance in RUU scenarios. However, insights should be integrated both from accompanying empirical work into comparative human performance across these same problem-question pairings, but also by integrating principles from Decision-Making under Deep Uncertainty (DMDU) approaches to effectively matriculate and address deep uncertainty related limitations within the problem space (e.g., shifting focus to decision robustness; for a review of approaches see Marchau et al., 2019). Both of these aspects are essential to understanding potential risks, but also in the improvement of model performance (e.g., via integrated APIs and prompt engineering). This endeavour will require a concerted effort, however the potential benefits extend beyond risk minimisation, to potentially substantial *gains* in dealing with the complex nature of real-world decision-making in uncertain and dynamic environments - which characterise many of the most critical problems we face globally.

In terms of AI safety policy, the emphasis must be on learning to be dynamic and adaptive. In this manner, policy solutions to RUUT are emblematic of dealing with the broader risk class of structural and socio-technical risk factors. Adopting a structural approach that considers multiple levers of change is key to addressing the *wicked* problem characteristics inherent to determining effective AI safety and governance. Such an approach includes not only the development of multi-faceted and flexible policies but also the establishment of oversight mechanisms that are capable of evolving with the changing landscape of AI technology and its applications. This approach is in line with 'adaptive foresight' (Eriksson & Weber, 2008) and other responses to policy-making when faced with deep uncertainty, and gel with the broader perspectives on socio-technical and structural risks and the recognition of the need for diverse insights in research, development, and policy formulation (e.g., Lazar & Nelson, 2023). For example, the rapid identification of structural risks (e.g., via structural risk evaluations incorporating techniques like system mapping) is not only necessary at the point of AI development governance, but also for policy-makers to identify system vulnerabilities (and conversely, leverage points) for mitigating or reversing harms across the broader spectrum of a changing society.

Furthermore, much like the raising of awareness regarding AI tool RUU capabilities, the advantages of adopting this perspective in policy-making is not limited to effective risk identification and mitigation efforts, but the vulnerabilities identified in system mapping and adaptive policy efforts are themselves often potential points of leverage for producing outsized *positive gains*. By way of analogy, identifying a tipping point potential in the establishment of a harmful norm (e.g., naive model misuse) reveals not only the vulnerability that should be mitigated in the case of *misuse* (e.g., raising awareness of RUU errors), but also unveils a leverage point to lean into (e.g., via increasing transparency) when that norm can instead be turned positive (e.g., AI tool use for RUU does now outperform human counterparts). Against the backdrop of fast moving and temporally sensitive, non-linear problems, identifying and exploiting non-linear intervention points is not only desirable, but a necessity, both in AI safety policy, and beyond.

The solutions for RUUT - as emblematic of structural risks more broadly - must be interconnected, with ongoing initiatives that emphasise oversight and transparency in AI development and use. This interconnectedness ensures that the efforts to mitigate the risks associated with AI tool misuse in RUU scenarios are comprehensive and aligned with the broader goals of AI safety and responsible innovation. In this regard, the example proposition of increased emphasis on oversight and transparency underscores the necessity of incorporating multiple perspectives and external oversight in AI governance (see e.g., Gruetzemacher et al., 2023).

In conclusion, addressing the RUUT problem requires a holistic approach that intertwines awareness, investigation, intervention, and policy adaptation. As we navigate an increasingly uncertain future, the need to understand and mitigate the risks associated with AI in complex decision-making becomes not just a technological imperative but a fundamental requirement for responsible governance and effective policy-making. By adopting this multifaceted approach, we can ensure that AI is used safely and effectively, harnessing its capabilities to navigate the uncertainties of our world while safeguarding against its potential pitfalls.

**Acknowledgements**

Our thanks to David Mannheim, Ross Gruetzemacher, Kyle Killian, and Iyngkarran Kumar for their comments and suggestions on earlier forms of this manuscript. This work was sponsored by the Transformative Futures Institute.

Correspondence should be addressed to Toby D. Pilditch at: tdpilditch@gmail.com